\documentclass[%
 reprint,
 amsmath,amssymb,
 aps,
]{revtex4-1}
\usepackage{graphicx}
\usepackage{amssymb}
\usepackage{amsmath}
\usepackage{epstopdf}
\usepackage{graphicx}
\usepackage{dcolumn}
\usepackage{bm}
\usepackage{hyperref}

\begin{document}

\preprint{APS/123-QED}

\title{Cooperative  epidemic spreading  on a two-layered interconnected network}

\author{Xiang Wei$^{1,2}$}
\author{Xiaoqun~Wu$^{2,3}$}%
\altaffiliation[]{Email: xqwu@whu.edu.cn}
\author{Shihua~Chen$^{3}$}%
\author{Jun-an Lu$^{3}$}%
\author{Guanrong Chen$^{4}$}%
\affiliation{%
 $^{1}$School of Engineering,
Honghe University, Yunan 661100,
China
}%
\affiliation{%
 $^{2}$School of Mathematics and Statistics,
 Wuhan University, Hubei 430072, China
}%
\affiliation{%
 $^{3}$Computational Science Hubei Key Laboratory, Wuhan University, Hubei
430072, China
}%
\affiliation{%
 $^{4}$Department of Electronic Engineering, City University
of Hong Kong, Hong Kong, China
}%




\date{\today}

\begin{abstract}
This study is concerned with the dynamical
behaviors of epidemic spreading  over a two-layered interconnected
network. Three models   in  different   levels are proposed to describe  cooperative spreading processes over the interconnected network, wherein the  disease in one network can spread to the other.
Theoretical analysis is provided for each model to reveal   that the global epidemic threshold in the interconnected network is not larger than the epidemic thresholds for the  two isolated layered networks.
In particular, in an  interconnected homogenous network, detailed theoretical analysis is presented, which allows quick and accurate calculations of the global epidemic threshold.
  Moreover, in an  interconnected heterogeneous network with inter-layer correlation between node degrees,  it is found that the inter-layer correlation coefficient has little impact on  the epidemic threshold, but has  significant impact   on the total prevalence.
Simulations further verify the analytical results, showing  that cooperative
epidemic processes promote the spreading of  diseases.
\end{abstract}

\pacs{Valid PACS appear here}
\maketitle


\section{\label{intro}Introduction}

Epidemic spreading, as an important  dynamic process taking place on complex networks,  has stimulated wide interest in the past two  decades.    Many spreading processes  have been  studied  on isolated networks \cite{Pastor2001Epidemic,Newman2002Spread, Motter2002Range,Eames2004Monogamous,Liljeros2003Sexual,Reed2006A}.  However,  in the real-world,  an epidemic process may spread across different networks.   For example,  Avian influenza, such as 2009 H1N1 and 2013 H7N9, can  spread  from poultry to humans.
A natural extension of the study is to use an interacting network model to analyze real-world epidemic processes,  where a disease is able to  spread from one network to another.     Recent studies have shown that comprising  interconnected networks can   more accurately simulate real-world situations \cite{Saumell2012Epidemic,Dickison2012Epidemics,Yagan2013Conjoining,Domenico2013Mathematical,Zhao2014Multiple,Xu2015Synchronizability,Li2015Bifurcation}.     The  classical SIS  and SIR models were  studied on interconnected networks.   In \cite{Saumell2012Epidemic},   a heterogeneous mean-field approach was developed to calculate  conditions for the emergence of an endemic state,   which revealed that the global epidemic threshold of interconnected networks is smaller than the threshold of each individual network.  In \cite{Dickison2012Epidemics},  the SIR model was studied on interconnected networks, which showed   that for strongly interconnected networks, the endemic state    occurs or dies out simultaneously in each component network.
 Using generating
function and bond percolation theory,   multiple routes
transmitted epidemic process was modeled following the SIR model in multiplex networks   \cite{Zhao2014Multiple},  which revealed that an endemic state can emerge in multiplex networks
even if the  layers are well below their respective epidemic thresholds.
In \cite{Cozzo2013Contact}, perturbation theory was used to analyze epidemic thresholds of networks, which   revealed that the spectral radius of the whole network matrix is never smaller  than that of each submatrix. Thus, the epidemic threshold of the whole network is always not larger  than the thresholds of  the isolated networks.
It was found that  the epidemic threshold of a multiplex network is  governed by the layer that contains  the largest  eigenvalue of the contact matrix\cite{Cozzo2013Contact,Wang2013Effect}. In \cite{Sanz2014Dynamics}, a framework was proposed to describe the spreading dynamics of two diseases, it was found there are regions of the parameter space, in which  a disease¡¯s outbreak is conditioned to the prevalence of the other disease.

The  interplay between awareness and disease spread processes is  a key issue in studying epidemics.  With  the  interaction of two processes, such as the spreading of a disease  and the spreading of the information awareness to prevent
infection,  epidemic spreading promotes  wider  information awareness which can further   protract   infection.
 A conclusion is  that information awareness can suppress  epidemic spreading \cite{Clara2013Dynamical,Granell2014Competing}.
Therefore, epidemic spreading processes on  multiplex networks exhibit rich phase diagrams of intertwined  effects.
While the study of uncorrelated complex networks is a fundamental step for investigating many  complex systems, in a  realistic interconnected network, inter-layer degree  correlation is
expected to exist. For example, in a social network, a person with a large number of links in one
network layer  is likely to    have many links in other types of network layers that reflect different kinds of social relations, such as being a friendly person\cite{Lee2011Correlated}.

Recent works have shown that   correlated multiplexity is ubiquitous in the world trade system \cite{Barigozzi2010Multinetwork}, as well as in transportation network systems\cite{Parshani2011Inter,Xu2011Interconnecting}. Due to their impact on network robustness \cite{Parshani2011Inter,Buldyrev2011Interdependent} and percolation properties \cite{Nicosia2015Measuring,Lee2011Correlated},   multiplex networks has been extensively studied.
In \cite{Wang2014Asymmetrically},   the impact of  inter-layer correlations on  epidemic processes was studied regarding awareness in disease networks. With a degree correlation between double-layer random awareness networks   and disease spreading, there is a strong evidence that an epidemic can be suppressed  through  large-degree nodes \cite{Wang2014Asymmetrically}, thus  it is possible to effectively mitigate an epidemic disease by     information diffusion    via hub nodes with  high degree centrality.

The structure of a complex network can be characterized by its degree property, which represents the type of interaction between components (nodes).  In terms of  degree distribution, complex networks can  be classified into homogeneous and heterogeneous networks\cite{Pastor2001Epidemic,Tanaka2011Dynamical}.  Homogeneous networks, such as random graphs and small-world models,  possess the Poisson type of degree distributions, with most nodes  basically bearing the mean degree. Heterogeneous networks, such as scale-free networks,   exhibit a power-law degree distribution.    This kind of distribution implies
that a small portion of nodes   have  very large degrees compared to
the average degree of the network.
However,  degree distribution     gives information about the connectivity probability at the level of  a group of nodes but not individual nodes.
  Thus, to obtain more detailed description for individual nodes, it is necessary to propose  a model describing spreading dynamics   at the individual level.

 The aforementioned works focus on multiplex networks, in which the nodes in each layer are the same, and  each node in one layer is only connected to its   counterparts in the other layer. However, many complex systems contain  layers with different kinds of nodes,  such as homosexual and  heterosexual networks of sexual contacts, transportation networks which depend  on different layers such as air routes, railways and roads.  Furthermore,  in these complex systems, a node in one layer can connect with various nodes in the other layer.  Thus, the study of epidemic spreading  on interconnected networks, in which different layers have different numbers or types of nodes, is of more practical significance.

The present study    focuses on how the   structures of complex networks and the inter-layer connections determine the epidemic
thresholds of a two-layered (extendable to more)  interconnected network.
 Three dynamical  models are  proposed to describe cooperative spreading processes  on an interconnected  homogenous or heterogeneous   network.
 The first one is the spreading dynamics on the whole network   of two homogeneous layered networks.
 The second is the  spreading dynamics in groups  of nodes with identical degrees on a two-layered heterogeneous  network with inter-layer correlations.
  The last one  is the spreading dynamical model at the  individual level  on a two-layered heterogeneous network without inter-layer correlations.
    This last model assumes that
probabilities of nodes being infected  are independent random variables.  Theoretical analysis and numerical simulations are used to investigate the cooperative spreading processes.

 The rest of the paper is organized as follows. Three models and theoretical analysis are presented in Sec. \ref{model}.    Numerical simulations are presented to illustrate the  behaviors of cooperative spreading processes  in Sec. \ref{simulation}. Finally, conclusions are given in Sec. \ref{Conclusion} with some  discussions.

 \section{\label{model}Network modeling and preliminaries}
\begin{figure}
  \includegraphics[width=0.4\textwidth]{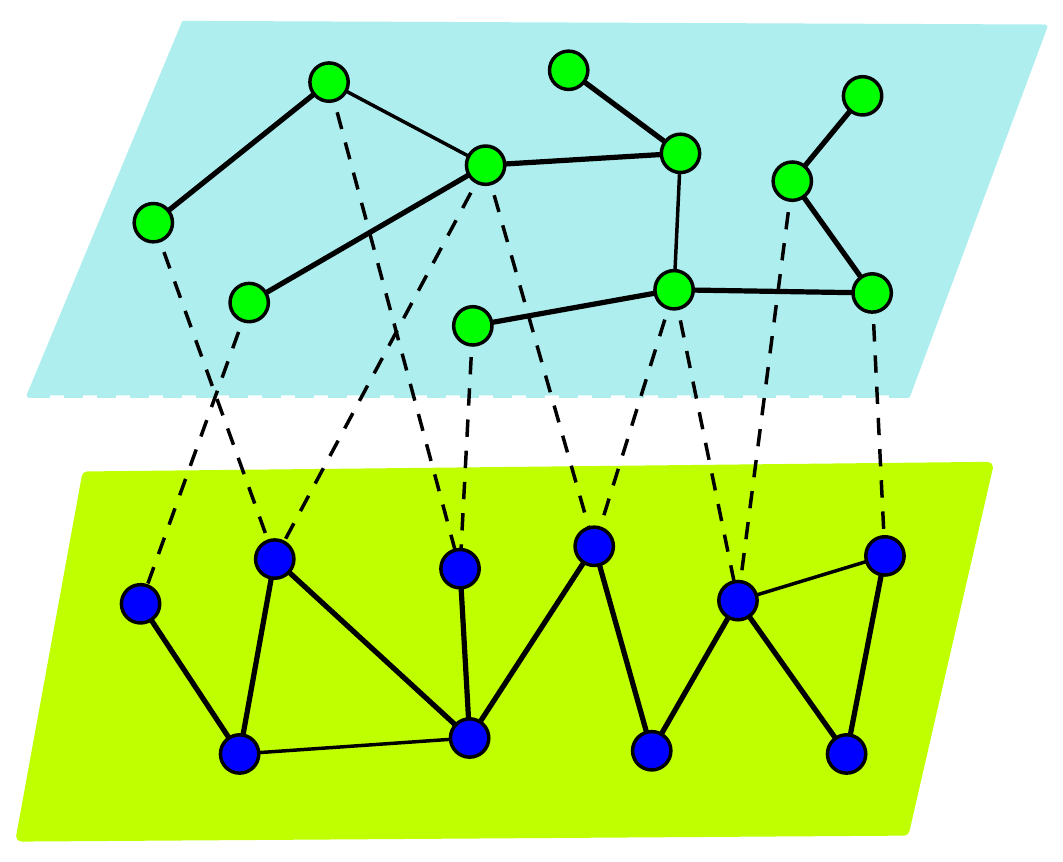}
\caption{The interconnected two-layer network is different from layer to layer, and  inter-layer connectivity randomly correlates between the two layers. Solid and dashed  lines  represent internal and inter-layer connections, respectively.}
\label{fig:model}       
\end{figure}

 Consider a two-layered interconnected  network, network layer
$A$ of size $N$ and  $B$ of size $M$, which have different connectivities.   Use network $AB$~$(BA)$, both of size $M+N$, to denote inter-layer connectivity from layer $A$~($B$) to $B$~($A$). The inter-layer connectivity randomly correlates between the two  layers.  An example of two interconnected  networks is shown in Fig.{\ref{fig:model}}.

The  classical susceptible-infected-susceptible (SIS) model \cite{Pastor2001Epidemic} is used to describe the spreading dynamics on the  interconnected network.     In the network, each node is   either in the   susceptible (S) or infected (I) state, and the  links represent the connections between nodes along which the infection can propagate.   At each time step, susceptible (S) nodes may be infected by  infected nodes within the same layer with a certain probability or in other layers with a certain probability  simultaneously.  On the other hand, infected nodes are recovered spontaneously with a certain probability on each layer. Let $\lambda_a$~$(\lambda_b)$ be the internal infection rate in layer $A~(B)$, and $\lambda_{ba}$~($\lambda_{ab}$) be the inter-layer infection rate from   nodes in $B$~($A$) to  nodes in $A$~($B$). Infected nodes are recovered  with   rate $\mu_a$~($\mu_b$) in $A~(B)$.
\subsection{A two-layered randomly-correlated homogeneous network }\label{model1}

{ Let  network layers $A$ and $B$ be
two homogeneous networks that are connected without relying on degree correlations. Nodes in layer
$A$ $(B)$ are characterized by average degree $\langle k_a\rangle $ ($\langle k_b\rangle $), and $\langle k_{ba}\rangle $ represents the average inter-layer   connections of nodes in $A$ and $\langle k_{ab}\rangle $  in $B$. The fractions of infected nodes in $A$ and $B$ are denoted by $\rho^A(t)$ and $\rho^B(t)$, respectively. Thus, the evolution processes can be written as}
\begin{align}\label{spread}
\frac{d\rho^A(t)}{dt}&=-\mu_a\rho^A(t)+\lambda_a\langle k_a\rangle \rho^A(t)(1-\rho^A(t))\notag \\
 &+\lambda_{ba}\langle k_{ba}\rangle \rho^B(t)(1-\rho^A(t)),\notag \\
\frac{d\rho^B(t)}{dt}&=-\mu_b\rho^B(t)+\lambda_b\langle k_b\rangle \rho^B(t)(1-\rho^B(t))\notag \\
&+\lambda_{ab}\langle k_{ab}\rangle \rho^A(t)(1-\rho^B(t)).
\end{align}
{In the first equation of Eq. (\ref{spread}), the first term on the right-hand side stands for the probability that nodes infected at time  $t$ are recovered; the second term is the probability that   susceptible nodes  are infected by their infected neighbors, which is  proportional to the infection rate $\lambda_a$ and the average degree $\langle k_a\rangle$; the last term represents the probability that   susceptible nodes in one layer  are infected by the infected neighbors from the other connected layer.  The second equation in Eq. (\ref{spread}) has similar meanings as the first one.

 \begin{figure}
  \includegraphics[width=0.4\textwidth]{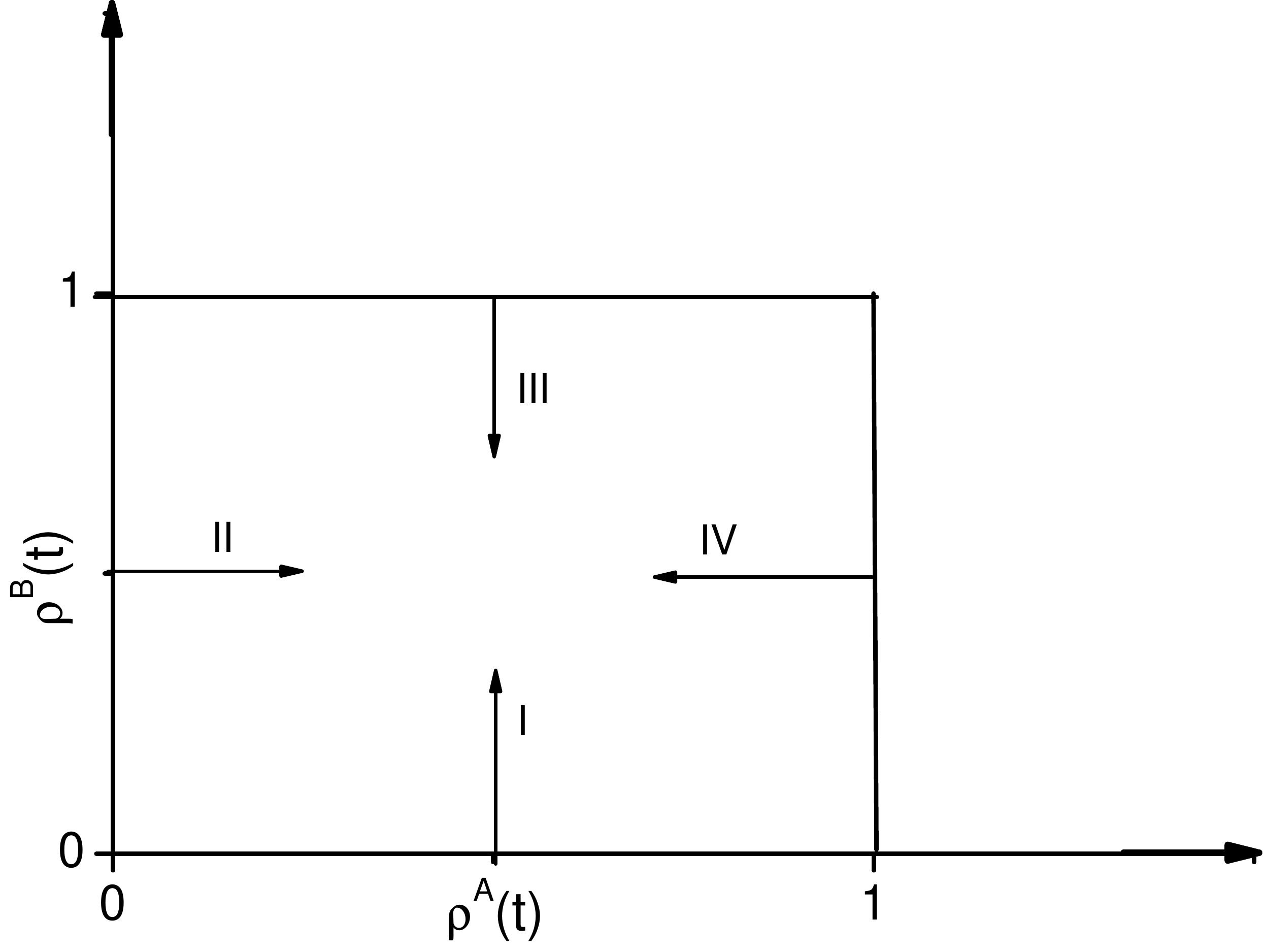}
\caption{ The movement  directions   of $\rho^A(t)$ and $\rho^B(t)$ from the boundaries. }
\label{trajectory}       
\end{figure}
 Without loss of generality, set $\mu_a=1$ and $\mu_b=1$.  After some transient time, the  dynamical system  (\ref{spread}) will evolve  into a stationary state.   To obtain  the nontrivial stationary solution of Eq. (\ref{spread}), one calculates
\begin{align}\label{spread_trivial}
0&=-\rho^A+\lambda_a\langle k_a\rangle \rho^A(1-\rho^A)+\lambda_{ba}\langle k_{ba}\rangle \rho^B(1-\rho^A)\notag, \\
0&=-\rho^B+\lambda_b\langle k_b\rangle \rho^B(1-\rho^B)+\lambda_{ab}\langle k_{ab}\rangle \rho^A(1-\rho^B).
\end{align}

 It can be seen that   a  global epidemic  activity will arise  in the  two connected layers if an epidemic disease  spreads in any layer.  The reason is  that the states $\rho^A\neq 0$ and $\rho^B= 0$~(or $\rho^A=0$ and $\rho^B\neq 0$) are not  equilibrium  of   Eq. (\ref{spread_trivial}).  Furthermore, since $\rho^A(t)\in [0,1]$ and $\rho^B(t)\in [0,1] $,   when $\rho^A(t)$ or $\rho^B(t)$ touches  the boundaries (except for the origin point), it will goes to some internal point in the region, as illustrated by   Fig.\ref{trajectory}.
 Particularly,   when $\rho^A(t)=0$ and $\rho^B(t) > 0$, one gets $\frac{d\rho^A(t)}{dt}>0$ from the first equation of Eq. (\ref{spread}). Thus, the trajectory of $\rho^A(t)$ will go upwards, as shown by the upward arrow  in the figure.   When $\rho^A(t)=1$, one obtains $\frac{d\rho^A(t)}{dt}<0$, thus $\rho^A(t)$ will go downwards. Analogously,   $\rho^B(t)$ will  go inwards if it touches the boundaries.

The critical point (global epidemic threshold) separating the healthy~($\rho^A=0$ and $\rho^B=0$) and endemic~($\rho^A\neq 0$ and $\rho^B\neq 0$) phases can be obtained by studying the stability of the absorbing solution of Eq. (\ref{spread}).  Suppose that the   global epidemic threshold  is  $\lambda^c$  for the two connected layers. Then   if $\lambda_a<\lambda^c$ and $\lambda_b<\lambda^c$, the system will go to the equilibrium $\rho^A=0$ and $\rho^B=0$;    otherwise, epidemic propagation through the inter-layer connectivity between   layers makes  the whole system evolve into an endemic state ($\rho^A\neq 0$ and $\rho^B\neq0$).  When  $\lambda_a$  or $\lambda_b$ approaches $ \lambda^c$, the prevalence of  infected nodes will appear:  $\rho^A\ll 1$ and $\rho^B\ll 1$.  Thus, around an endemic state, one can neglect  the second-order terms in  Eq. (\ref{spread_trivial}), so as to obtain
\begin{eqnarray}\label{conclusion}
\left[
\begin{array} {lcccr}
  \lambda_a\langle k_a\rangle -1&\lambda_{ba}\langle k_{ba}\rangle \\
  \lambda_{ab}\langle k_{ab}\rangle &\lambda_b\langle k_b\rangle -1\\
\end{array}
\right]
\left[
\begin{array} {lcccr}
  \rho^A\\
  \rho^B\\
\end{array}
\right]=0.
\end{eqnarray}
One can write Eq. (\ref{conclusion}) as
\begin{eqnarray}\label{det}
\frac{d\rho}{dt}=-\rho+C\rho,
\end{eqnarray}
where   $\rho=(\rho^A,\rho^B)^\top$ and
\begin{eqnarray}\label{C_matrix}
C=
\left[
\begin{array} {lcccr}
  \lambda_a\langle k_a\rangle &\lambda_{ba}\langle k_{ba}\rangle \\
  \lambda_{ab}\langle k_{ab}\rangle &\lambda_b\langle k_b\rangle \\
\end{array}
\right].
\end{eqnarray}
{It is obvious that the global endemic state will not  arise whenever the maximum eigenvalue of matrix $C$ satisfies $\Lambda_{\mbox{max}}(C)<1$, but if  $\Lambda_{\mbox{max}}(C)>1$, a global endemic state will  arise in the interconnected network. Thus, the critical epidemic point  (threshold) is determined by $\Lambda_{\mbox{max}}(C)=1$.}

{When  $1$ is the eigenvalue of $C$, it means $\lambda_a$  or $\lambda_b$ approaches $\lambda^c$. By replacing $\lambda_a$ and $\lambda_b$ with the threshold $\lambda^c$, one can calculate the threshold from the following equation:}
\begin{eqnarray}\label{det0}
&(\lambda^c)^2\langle k_a\rangle \langle k_b\rangle-\lambda^c(\langle k_a\rangle+\langle k_b\rangle)\notag \\
&+1-\lambda_{ab}\lambda_{ba}\langle k_{ab}\rangle\langle k_{ba}\rangle=0.
\end{eqnarray}
Eq. (\ref{det0}) has two solutions. By using the  smaller one as the global epidemic threshold\cite{Clara2013Dynamical}, one obtains
\begin{eqnarray}\label{Seeking_roots}
\lambda^c=\frac{\langle k_a\rangle+\langle k_b\rangle-\sqrt{(\langle k_a\rangle+\langle k_b\rangle)^2-f}}{2\langle k_a\rangle\langle k_b\rangle},
\end{eqnarray}
where $f=4\langle k_a\rangle \langle k_b\rangle (1-\lambda_{ab}\lambda_{ba}\langle k_{ab}\rangle\langle k_{ba}\rangle)>0$.

In addition, for the isolated layer $A$,  the epidemic  process  is described by
\begin{align}\label{isolate_spread}
\frac{d\rho^A(t)}{dt}&=-\rho^A(t)+\lambda_a\langle k_a\rangle \rho^A(t)(1-\rho^A(t)).
\end{align}
Similarly, by neglecting second-order terms, one obtains
 \begin{align}\label{spread5}
0=(1-\lambda_a\langle k_a\rangle) \rho^A.
\end{align}
Thus, one obtains the epidemic  threshold of isolated  layer $A$ as $\lambda_a^c=\frac{1}{\langle k_a\rangle}$. Similarly, one can obtain the   threshold $\lambda_b^c=\frac{1}{\langle k_b\rangle}$ for the isolated layer $B$. Analogously, one can obtain the epidemic threshold $\lambda_{ab}^c=\frac{1}{\langle k_{ab}\rangle}$ for inter-layer connections from layer $A$  to $B$, and  $\lambda_{ba}^c=\frac{1}{\langle k_{ba}\rangle}$  from layer $B$  to $A$.

 In order to compare the global epidemic threshold of the interconnected network with  those of the corresponding isolated networks,   assume  that neither the inter-layer network $AB$ nor network $BA$ is already in an endemic state, which means that $\lambda_{ab}<\lambda^c_{ab}$ and $\lambda_{ba}<\lambda^c_{ba}$, thus, $\lambda_{ab} \langle k_{ab}\rangle<1$ and $\lambda_{ba}\langle k_{ba}\rangle<1$.

When $\langle k_a\rangle>\langle k_b\rangle$, meaning  $\lambda_a^c<\lambda_b^c$, one obtains
\begin{eqnarray} \label{global_threshold1}
\lambda^c<\frac{\langle k_a\rangle+\langle k_b\rangle-(\langle k_a\rangle-\langle k_b\rangle)}{2\langle k_a\rangle\langle k_b\rangle}=\frac{1}{\langle k_a\rangle}=\lambda_a^c<\lambda_b^c.
\end{eqnarray}
When $\langle k_b\rangle>\langle k_a\rangle$, which means  $\lambda_b^c<\lambda_a^c$, one obtains
\begin{eqnarray} \label{global_threshold2}
\lambda^c<\frac{\langle k_a\rangle+\langle k_b\rangle-(\langle k_b\rangle-\langle k_a\rangle)}{2\langle k_a\rangle\langle k_b\rangle}=\frac{1}{\langle k_b\rangle}=\lambda_b^c<\lambda_a^c.
\end{eqnarray}
When $\langle k_b\rangle=\langle k_a\rangle$, meaning $\lambda_b^c=\lambda_a^c$, one still obtains
\begin{eqnarray} \label{global_threshold3}
\lambda^c<\frac{\langle k_a\rangle+\langle k_b\rangle}{2\langle k_a\rangle\langle k_b\rangle}=\frac{1}{\langle k_a\rangle}=\lambda_a^c=\lambda_b^c.
\end{eqnarray}
Therefore, the conclusion is that the global epidemic threshold of the  interconnected network is smaller
than the epidemic thresholds of the  isolated layers. This implies that cooperative spreading  promotes   epidemic processes.
Furthermore, Eq. (\ref{Seeking_roots}) presents an analytical formula for the global epidemic threshold in  the interconnected network based on macroscopic information within each layer and across layers, such as the average degrees and inter-layer infection rates.

 The preceding conclusion reveals that there are some situations where neither isolated layers $A$ and $B$  nor  networks $AB$ and  $BA$ are in endemic state yet the interconnected network evolves into an endemic state.  This interesting case  is investigated to find under what condition an endemic state exists in the  interconnected network. 


    From Eq.(\ref{C_matrix}), by denoting
\begin{eqnarray}\label{C_matrix1}
C=
\left[
\begin{array} {lcccr}
  \lambda_a\langle k_a\rangle &\lambda_{ba}\langle k_{ba}\rangle \\
  \lambda_{ab}\langle k_{ab}\rangle &\lambda_b\langle k_b\rangle \\
\end{array}
\right]
=
\left[
\begin{array} {lcccr}
  \delta_a & \delta_{ba} \\
  \delta_{ab} &\delta_b \\
\end{array}
\right],
\end{eqnarray}
one obtains
\begin{eqnarray}\label{det}
\Lambda_{\mbox{max}}=\frac{\delta_a+\delta_b+\sqrt{(\delta_a-\delta_b)^2+4\delta_{ab}\delta_{ba}}}{2}.
\end{eqnarray}
Note that in this  case, each element of matrix $C$  is in   the range $(0,1)$. { In the simulation section,    phase diagrams for $\Lambda_{\mbox{max}}$ will be numerically display (i.e., the maximum eigenvalue of matrix $C$)  in Figs. \ref{Eigenvalues_greater_than_one1}, \ref{Eigenvalues_greater_than_one2} and \ref{Eigenvalues_greater_than_one3}    with respect to parameters $\delta_a$ and   $\delta_{ab}$ for three specific cases.}

Further, in some specific  scenarios, one can easily derive  the global epidemic threshold from the epidemic thresholds of isolated layer networks.


(1)~When the inter-layer infection rates  are much smaller  than the intra-layer infection rates (e.g., $\lambda_{ab}\ll \lambda_a$)   and  the inter-layer  average degrees are smaller than the intra-layer ones (i.e., $\langle k_{ab}\rangle <\langle k_{a}\rangle$ and $\langle k_{ba}\rangle<\langle k_{b}\rangle$), it follows from Eq. (\ref{det0}) that
\begin{eqnarray}\label{E15}
(\lambda^c\langle k_a\rangle-1)(\lambda^c\langle k_b\rangle-1)\approx 0.
\end{eqnarray}
Therefore,  the epidemic threshold $\lambda^c=\frac 1{k_{max}}$, where $k_{max}$ is the larger value between $\langle k_a\rangle$ and $\langle k_b\rangle$. This means that when there is weak inter-layer infection between layers,  the layer with a smaller epidemic threshold dominates  the global epidemic threshold of the interconnected network.

(2)~When the  inter-layer  infection rates are equal to the intra-layer ones, and  the inter-layer  average degrees are equal to the intra-layer ones,  i.e., $\lambda_{ab}\approx \lambda_a$ and $\lambda_{ba}\approx \lambda_a$,  and $\langle k_{ab}\rangle \approx \langle k_{a}\rangle$, $\langle k_{ba}\rangle\approx\langle k_{b}\rangle$), one has   $\lambda_a\lambda_b\langle k_a\rangle\langle k_b\rangle\approx \lambda_{ab}\lambda_{ba}\langle k_{ab}\rangle\langle k_{ba}\rangle$. Therefore, from   Eq. (\ref{det0}), one obtains
\begin{eqnarray}\label{}
1-\lambda^c\langle k_a\rangle-\lambda^c\langle k_b\rangle\approx 0.
\end{eqnarray}
  Thus, $\lambda^c=\frac 1{k_{a}+k_{b}}$.

(3) When the  inter-layer  infection rates are much larger  than the intra-layer ones, and  the inter-layer  average degrees are larger than the intra-layer ones, i.e., $\lambda_{ab}\gg \lambda_a$ and $\lambda_{ba}\gg \lambda_a$,     $\langle k_{ab}\rangle > \langle k_{a}\rangle$ and $\langle k_{ba}\rangle>\langle k_{b}\rangle$, the inter-layer infection rates will be dominant in determining  the epidemic process.   Therefore, from   Eq. (\ref{det0}), after using $\lambda^c$ to substitute $\lambda^{ab}$ and $\lambda^{ba}$, one obtains
\begin{eqnarray}\label{E17}
1-\lambda^{c}\langle k_{ba}\rangle\lambda^{c}\langle k_{ab}\rangle\approx 0.
\end{eqnarray}
Thus, $\lambda^c=\frac 1{\sqrt{\langle k_{ab}\rangle \langle k_{ba}\rangle }}$.

Summarizing, the epidemic threshold of the interconnected network  mainly depends on the average  inter-layer   or intra-layer degrees, whichever is larger.
 When a disease begins to spread, it is easy to lead to an earlier endemic activity in a more densely populated region because it has a lower epidemic  threshold, and then the disease will spread to other connected regions, which will eventually result in a global endemic state in  the whole population. Therefore, to inhibit the spreading of a disease, an effective measure is to decrease the population density, as is intuitively clear.  }

\subsection{A two-layered   correlated heterogeneous network }

Consider epidemic processes over a two-layered heterogeneous network.  Assume that all nodes of the same   degree behave equally.  Define the partial prevalence   $\rho_{k_a}^A(t)$  ($\rho_{k_b}^B(t)$) as the fraction of infected nodes with a given degree $k_a$ ($k_b$)  in layer $A~(B)$.     The goal is to understand the impact of the correlation  of inter-layer connectivity structures on the epidemic processes, such as   epidemic thresholds and    prevalence.
 In   the interconnected heterogeneous network, let $P(k)$ denote the probability that a node has degree $k$ within a network layer,  and  $P(k^{'}|k)$ be   the conditional probability  that a node of degree $k$ in one layer is connected to a node of degree $k^{'}$ in the other layer.    The normalization conditions $\sum_{k}P(k)=1$ and $\sum_{k}P(k^{'}|k)=1$
hold.  Thus, the average number of
links connecting a node of degree $k$ to some nodes of degree $k^{'}$ is $kP(k^{'}|k)$. For simplicity, only consider the   degree correlation of  inter-layer connectivity but not that   of internal layers.

   By Eq. (\ref{spread}), the evolution processes  can be written as
\begin{align}\label{mesoscale_spread}
 \frac{d\rho_{k_{a}}^A(t)}{d t}&=-\rho_{k_{a}}^A(t)+\lambda_ak_{a}(1-\rho_{k_{a}}^A(t))\Theta_{k_a}^A(t)\notag \\
 &+\lambda_{ba}k_{ba}(1-\rho^A_{k_{a}}(t))\Theta_{k_b}^{BA}(t)\notag, \\
\frac{d\rho_{k_{b}}^B(t)}{d t}&=-\rho_{k_{b}}^B(t)+\lambda_bk_{b}(1-\rho_{k_{b}}^B(t))\Theta_{k_b}^B(t)\notag \\&+\lambda_{ab}k_{ab}(1-\rho_{k_{b}}^B(t))\Theta_{k_a}^{AB}(t),
\end{align}
where
\\$\Theta_k^A(t)=\frac{1}{\langle k_{a}\rangle}\sum_{k_{a}^{'}}k_{a}^{'}P(k_{a}^{'})\rho^A_{k_{a}^{'}}(t)$,
\\ $\Theta_k^B(t)=\frac{1}{\langle k_{b}\rangle}\sum_{k_{b}^{'}}k_{b}^{'}P(k_{b}^{'})\rho^B_{k_{b}^{'}}(t)$,
\\$\Theta_k^{BA}(t)=\sum_{k_{b}^{'}}P(k_{b}^{'}|k_{a})\rho^B_{k_{b}^{'}}(t)$,
\\$\Theta_k^{AB}(t)=\sum_{k_{a}^{'}}P(k_{a}^{'}|k_{b})\rho^A_{k_{a}^{'}}(t)$.\\
 In the first equation of Eq. (\ref{mesoscale_spread}), the first term   on the right-hand side means   that infected  nodes of degree $k_a$ in layer $A$ can be recovered.  The second term means that susceptible nodes of degree $k_a$ are infected by their infected neighbors within the same layer,  where  $1-\rho_{k_{a}}^A(t)$ represents  the fraction of  susceptible nodes of degree $k_a$, $\Theta_{k_a}^A(t)$  is the probability that a link emanating from the nodes of degree $k_a$ points to an   infected node within  layer $A$. The last term appears due to the coupling of layer $A$ with layer $B$,  and stands for a similar function as that of the second term,  except that the variable $\Theta_k^{BA}(t)$ is the probability that a link emanating from the nodes of degree $k_a$ points to an  inter-layer infected node. The second equation is analogous.

{$\mathbf{Lemma~1}$\cite{Bapat2010Graphs} $(\mathbf{Cauchy~interlacing ~theorem})$ Let $A$ be a symmetric $n\times n$ matrix and let $B$ be a principal submatrix of $A$ of order $n-1$. If $\beta_1\geq \beta_2\geq \cdots \geq \beta_n$ and $\gamma_1\geq \gamma_2\geq \ldots \geq\gamma_{n-1}$ are the eigenvalues of $A$ and $B$ respectively, then}
\begin{align}\label{value1}
\beta_1\geq\gamma_1\geq \beta_2\geq\cdots\beta_{n-1}\geq\gamma_{n-1}\geq\beta_n.
\end{align}
  To calculate the stationary solution of Eq. (\ref{mesoscale_spread}),  let
\begin{align}\label{mesoscale_spread_final}\notag \\
&0=-\rho_{k_{a}}^A+\lambda_ak_{a}\frac{1}{\langle k_{a}\rangle}\sum_{k_{a}^{'}}k_{a}^{'}P(k_{a}^{'})\rho^A_{k_{a}^{'}}\notag \\
&+\lambda_{ba}k_{ba}\sum_{k_{b}^{'}}P(k_{b}^{'}|k_{a})\rho^B_{k_{b}^{'}},\notag \\
&0=-\rho_{k_{b}}^B+\lambda_bk_{b}\frac{1}{\langle k_{b}\rangle}\sum_{k_{b}^{'}}k_{b}^{'}P(k_{b}^{'})\rho^B_{k_{b}^{'}}\notag\\
&+\lambda_{ab}k_{ab}\sum_{k_{a}^{'}}P(k_{a}^{'}|k_{b})\rho^A_{k_{a}^{'}}.
\end{align}

For the two-layered interconnected network, similarly to the analysis in Subsec.\ref{model1}, there exists a critical point  separating a healthy phase with $\rho^A_{k_{a}}=\rho^B_{k_{b}}=0$ and an endemic phase  with $\rho^A_{k_{a}}\neq 0$ and $\rho^B_{k_{b}}\neq 0$.    Analogously, by neglecting the second-order terms in  Eq. (\ref{mesoscale_spread_final}) around $\rho^A_{k_{a}}=\rho^B_{k_{b}}=0$ and replacing $\lambda_a$ and $\lambda_b$ with the epidemic threshold $\lambda^c$,  one obtains
\begin{align}\label{correlation}
\left[
\begin{array} {lcccr}
  &C^A&C^{BA}\\
  &C^{AB}&C^B\\
\end{array}
\right]
*\left[
\begin{array} {lcccr}
  \rho^A\\
  \rho^B\\
\end{array}
\right]
-\frac{1}{\lambda^c}I\left[
\begin{array} {lcccr}
  \rho^A\\
  \rho^B\\
\end{array}
\right]=0,
\end{align}
where $\rho^A=[\rho_{k_1}^A,\rho_{k_2}^A,...,\rho_{k_{l_1}}^A]^\top$, $l_1$ represents the distinct node degrees of nodes in layer $A$, and $\rho^B=[\rho_{k_1}^B,\rho_{k_2}^B,...,\rho_{k_{l_2}}^B]^\top$, $l_2$ stands for the distinct node degrees   in layer $B$, $I$ is the identity matrix, and
\\$C^A(k_a,k_{a}^{'})=k_{a}k_{a}^{'}P(k_{a}^{'})/\langle k_{a}\rangle$,
\\$C^{BA}(k_a,k_{b}^{'})=\lambda_{ba}k_{ba}P(k_{b}^{'}|k_{a})$,
\\$C^{AB}(k_b,k_{a}^{'})=\lambda_{ab}k_{ab}P(k_{a}^{'}|k_{b})$,
\\$C^B(k_b,k_{b}^{'})= k_{b}k_{b}^{'}P(k_{b}^{'})/\langle k_{b}\rangle$,\\$k_a=k_1,k_2,...,k_{l_1}, k_a^{'}=k_1,k_2,...,k_{l_1}$,
\\$k_b=k_1,k_2,...,k_{l_2}, k_{b}^{'}=k_1,k_2,...,k_{l_2}$.

 Denote
\begin{align}\label{}
L=\left[
\begin{array} {lcccr}
C^A&C^{BA}\\
C^{AB}&C^B\\
\end{array}
\right],
\end{align}
which is named the supra-connectivity matrix.
For an   isolated layer $A$ without interconnection with external networks, the epidemic dynamics is
\begin{align}\label{mesoscale_spread_isolate}
 \frac{d\rho_{k_{a}}^{A}(t)}{dt}&=\lambda_ak_{a}(1-\rho_{k_{a}}^{A}(t))\frac{1}{\langle k_{a}\rangle}\sum_{k_{a}^{'}}k_{a}^{'}P(k_{a}^{'})\rho^{A}_{k_{a}^{'}}(t)\notag\\
 &-\rho_{k_{a}}^{A}(t).
\end{align}
Similarly, by calculating the epidemic threshold of the isolated layer from Eq. (\ref{mesoscale_spread_isolate}),   one obtains
\begin{align}\label{mesoscale_spread_isolate1}
\Bigg[C^A-\frac{1}{\lambda_a}I\Bigg]\rho^{A}=0,
\end{align}
which  has a nonzero solution ($\rho^{A}> 0$)  if and only if $1/\lambda_a$ is an eigenvalue of matrix $C^A$.
Similarly, for   isolated layer $B$, one has \begin{align}\label{mesoscale_spread_isolate1}
\Bigg[C^B-\frac{1}{\lambda_b}I\Bigg]\rho^{B}=0,
\end{align}
which  has a nonzero solution ($\rho^{B}> 0$)  if and only if $1/\lambda_b$ is an eigenvalue of matrix $C^B$.
Thus, the epidemic thresholds for isolated layer $A$ and $B$ are determined by the maximum eigenvalues of  $C^A$ and $C^B$, respectively. While   for the two interconnected layers $A$ and $B$,  the epidemic threshold is determined by  the maximum  eigenvalue of $L$. According to Lemma~1, since $C^A$ and $C^B$ are both sub-matrices of $L$, one has
$ \Lambda_{\mbox{max}}(L)  \geq  \Lambda_{\mbox{max}}(C^A) $ and $ \Lambda_{\mbox{max}}(L)  \geq \Lambda_{\mbox{max}}(C^B) $, where $\Lambda_{\mbox{max}}(R)$ represents the maximum eigenvalue of matrix $R$.

Therefore,
$\lambda^c=\frac{1}{\Lambda_{\mbox{max}}(L)}\leq \frac{1}{\Lambda_{\mbox{max}}(C^A)}=\lambda_A$ and $\lambda^c=\frac{1}{\Lambda_{\mbox{max}}(L)}\leq \frac{1}{\Lambda_{\mbox{max}}(C^B)}=\lambda_b$. That is,   the global epidemic threshold of the interconnected network is not larger than the epidemic thresholds of the corresponding isolated networks.

In  numerical simulations below,  the impact of inter-layer correlation of nodes with different degrees on the global epidemic threshold and {  total prevalence  will be analyzed, which is  defined as}
 \begin{align}\label{eq:rho26}
\rho=\frac{1}{l_1}\sum_{j=1}^{l_1}\rho^A_{k_{l_1}}+\frac{1}{l_2}\sum_{j=1}^{l_2}\rho^B_{k_{l_2}}.
\end{align}

\subsection{An uncorrelated two-layered   heterogeneous  network}
\label{sec:1}
In this subsection,  the epidemic processes over two  interconnected heterogeneous networks $A$ and $B$  will  be investigated at the level of individual nodes.
 {For simplicity, denote the adjacency matrix of network $A$ as $ A=(a_{ij})_{N\times N}$, that for network $B$ be $ B=(b_{ij})_{M\times M}$, that for the  external network $AB$ from network $A$  to $B$ be $ AB=C =(c_{ij})_{N\times M}$, and  that for the external network  $BA$ from $B$  to $A$ be $ BA=D=(d_{ij})_{M\times N}$. Specifically,  if there is a link from node $j$ to node
$i~( j\neq i) $ in layer $A$,  then  $a_{ij}=1$; otherwise, $a_{ij}=0$. Similarly for $B, C$ and $D$.}


      Let $\rho_i^A(t)~(\rho_i^B(t))$ stand for the probability that an individual node $i$ is infected at time $t$ in layer $A~(B)$.  Then, the   evolution of the probability of infection of any node $i$ reads
\begin{align}\label{spread_micro}
 \rho_i^A(t+1)&=(1-\mu_a)\rho_i^A(t)+(1-\rho_i^A(t))(1-q_i^A(t))\notag \\
 &+(1-\rho_i^A(t))(1-q_i^{BA}(t)),~i=1,2,...,N,\notag \\
\rho_i^B(t+1)&=(1-\mu_b)\rho_i^B(t)+(1-\rho_i^B(t))(1-q_i^B(t))\notag \\
 &+(1-\rho_i^B(t))(1-q_i^{AB}(t)),~i=1,2,...,M,
\end{align}
where $\mu_a$~($\mu_b$) is the the recovery rate of the infected nodes in layer $A$~($B$), $q_i^A(t)$~($q_i^B(t)$) is the probability of node $i$ not being infected by any internal neighbor in layer $A$~($B$), and $q_i^{AB}(t)$~($q_i^{BA}(t)$) is the probability of node $i$ in layer $A$ ($B$) not being infected by any inter-layer neighbor in $B$~($A$).  In detail,
\begin{eqnarray}\label{nospread}
&q^A_{i}(t)=\prod_{j=1}^{N}(1-\lambda_aa_{ij}\rho^A_{j}(t)),\notag \\
&q^B_{i}(t)=\prod_{j=1}^{M}(1-\lambda_bb_{ij}\rho^B_{j}(t)),\notag \\
&q^{AB}_{i}(t)=\prod_{j=1}^{M}(1-\lambda_{ba}c_{ij}\rho^A_{j}(t)),\notag \\
&q^{BA}_{i}(t)=\prod_{j=1}^{N}(1-\lambda_{ab}d_{ij}\rho^B_{j}(t)).
\end{eqnarray}

 In the first equation in Eq. (\ref{spread_micro}), the first term on the right-hand stands for the probability that node $i$ is infected at time $t$ but is not recovered,  the second term  is the probability that susceptible  node $i$   is infected by at least one internal neighbor, and the last term is a similar function  as the second term except that the node is infected by at least one inter-layer infected neighbor.  The second equation in Eq. (\ref{spread_micro}) has the same meaning as the first one.

Similarly to the analysis in  Subsection \ref{model1}, there exists a global epidemic threshold $\lambda^c$ for the two-layered interconnected heterogeneous  network. When $\lambda_a$  or $\lambda_b$ approaches $\lambda^c$, the probabilities satisfy $0<\rho_i^A\ll 1$ and $0<\rho_i^B\ll 1$. Thus, by neglecting the second-order terms in
Eq.~(\ref{nospread}), one obtains
\begin{eqnarray}\label{newnospread}
q_{i}^A(t)\approx 1- \lambda_a\sum_{j=1}^{N}a_{ij}\rho^A_{j}(t),\notag \\
q_{i}^B(t)\approx 1-\lambda_b\sum_{j=1}^{M}b_{ij}\rho^B_{j}(t),\notag \\
q_{i}^{BA}(t)\approx 1- \lambda_{ba}\sum_{j=1}^{N}c_{ij}\rho^B_{j}(t),\notag\\
q_{i}^{AB}(t)\approx 1-\lambda_{ab}\sum_{j=1}^{M}d_{ij}\rho^A_{j}(t).
\end{eqnarray}
 By substituting  Eq. (\ref{newnospread}) into Eq. (\ref{spread_micro}), one  obtains
\begin{align}\label{spread_micro1}
\rho_i^A(t+1)&=(1-\mu_a)\rho_i^A(t)+\lambda_a(1-\rho_i^A(t))\sum_{j=1}^{N}a_{ij}\rho_j^A(t)\notag \\
 &+\lambda_{ba}(1-\rho_i^A(t))\sum_{j=1}^{M}c_{ij}\rho_j^B(t),~i=1,2,...,N, \notag\\
\rho_i^B(t+1)&=(1-\mu_b)\rho_i^B(t)+\lambda_b(1-\rho_i^B(t))\sum_{j=1}^{M}b_{ij}\rho_j^B(t)\notag \\
 &+(1-\rho_i^B(t))\sum_{j=1}^{N}d_{ij}\rho_j^A(t),~i=1,2,...,M.
\end{align}
By neglecting second-order terms, one can easily calculate the nontrivial stationary solution of Eq. (\ref{spread_micro1}) by the fixed point iteration method  as follows:
\begin{eqnarray}\label{spread_micro2}
\rho_i^A=(1-\mu_a)\rho_i^A+\lambda_a\sum_{j=1}^{N}a_{ij}\rho_j^A\notag
 +\lambda_{ba}\sum_{j=1}^{M}c_{ij}\rho_j^B, \notag\\
\rho_i^B=(1-\mu_b)\rho_i^B+\lambda_b\sum_{j=1}^{M}b_{ij}\rho_j^B
 +\lambda_{ab}\sum_{j=1}^{N}d_{ij}\rho_j^A.
\end{eqnarray}
After substituting $\lambda_a$ and $\lambda_b$ with the global epidemic threshold $\lambda^c$,  one obtains
\begin{eqnarray}\label{matrix11}
A\rho^{A}-\frac{\mu_a}{\lambda^c}I_N\rho^{A}+\frac{\lambda_{ba}}{\lambda^c}C\rho^{B}=0,\notag \\
B\rho^{b}-\frac{\mu_b}{\lambda^c}I_M\rho^{B}+\frac{\lambda_{ab}}{\lambda^c}D\rho^{A}=0,
\end{eqnarray}
where $\rho^A=[\rho_{1}^A,\rho_{2}^A,...,\rho_{N}^A]^\top$ and $\rho^B=[\rho_{1}^B,\rho_{2}^B,...,\rho_{M}^B]^\top$,  and  the total prevalence  for the   interconnected networks is defined  as
 \begin{align}\label{prevalence1}
\rho=\frac{1}{N}\sum_{j=1}^N\rho^A_{j}+\frac{1}{M}\sum_{j=1}^M\rho^B_{j}.
\end{align}

From Eq. (\ref{matrix11}), one has
\begin{align}\label{sustain}
\left[
\begin{array} {lcccr}
  &A&\frac{\lambda_{ba}}{\lambda^c}C\\
  &\frac{\lambda_{ab}}{\lambda^c}D&B\\
\end{array}
\right]
*\left[
\begin{array} {lcccr}
  \rho^A\\
  \rho^B\\
\end{array}
\right]
\\-\left[
\begin{array} {lcccr}
  &\frac{\mu_a}{\lambda^c}I_N& 0\\
  &0&\frac{\mu_b}{\lambda^c}I_M\\
\end{array}
\right]
*\left[
\begin{array} {lcccr}
  \rho^A\\
  \rho^B\\
\end{array}
\right]
=0.
\end{align}
Denote
\begin{align}\label{}
L=\left[
\begin{array} {lcccr}
  &A&\frac{\lambda_{ba}}{\lambda^c}C\\
  &\frac{\lambda_{ab}}{\lambda^c}D&B\\
\end{array}
\right],
\end{align}
and  call it the supra-adjacency matrix.

For comparison, consider   two  isolated layers $A$ and $B$,  with probabilities of infection being denoted by $\rho^A_{i}(t)$ and $\rho^B_{i}(t)$ for node $i$:
\begin{eqnarray}\label{isolate_spread}
\rho_i^A(t+1)&=(1-\mu_a)\rho_i^A(t)+(1-\rho_i^A(t))(1-q_i^A(t))\notag,\\
\rho_i^B(t+1)&=(1-\mu_b)\rho_i^B(t)+(1-\rho_i^B(t))(1-q_i^B(t)).
\end{eqnarray}

 By substituting Eq. (\ref{nospread}) into Eq. (\ref{isolate_spread}) and neglecting the
second-order terms, one can calculate  the nontrivial stationary solution for isolated layers  as follows,
\begin{eqnarray}\label{isotatematrix}
\Bigg(A-\frac{\mu_a}{\lambda_a}I\Bigg)\rho^{A*}=0,\notag \\
\Bigg(B-\frac{\mu_b}{\lambda_b}I\Bigg)\rho^{B*}=0,
\end{eqnarray}
where $\rho^A=[\rho_{1}^{A*},\rho_{2}^{A*},...,\rho_{N}^{A*}]^\top$, $\rho^B=[\rho_{1}^{B*},\rho_{2}^{B*},...,\rho_{M}^{B*}]^\top$. The total prevalence  for the two isolated layers $A$ and $B$ is defined as
 \begin{align}\label{prevalence2}
\rho^*=\frac{1}{N}\sum_{j=1}^N\rho^{A*}_{j}+\frac{1}{M}\sum_{j=1}^M\rho^{B*}_{j}.
\end{align}
Eq. (\ref{isotatematrix}) has a nontrivial solution if and only if
$\frac{\mu_a}{\lambda_a} $ and $\frac{\mu_b}{\lambda_b}$ are  eigenvalues of matrix $A$ and $B$, respectively, that is,
\begin{align}\label{value1}
\lambda_a^*=\frac{\mu_a}{\Lambda_{\mbox{max}} (A)},\notag \\
\lambda_b^*=\frac{\mu_b}{\Lambda_{\mbox{max}} (B)}.
\end{align}
 One can compare  the global epidemic threshold of the interconnected network with those of the corresponding isolated layers for the following two scenarios:\\
(1) The case of $\mu_a=\mu_b$\\
Eq.~(\ref{sustain}) has nontrivial solutions if and only if $\frac{\mu_a}{\lambda^c}$ is an eigenvalue of $L$. According to Lemma~1, since $A$ and $B$ are both sub-matrices of $L$, one has
$ \Lambda_{\mbox{max}}(L)  \geq  \Lambda_{\mbox{max}}(A) $ and $ \Lambda_{\mbox{max}}(L)  \geq \Lambda_{\mbox{max}}(B) $, where $\Lambda_{\mbox{max}}(R)$ represents the maximum eigenvalue of matrix $R$.

Thus,    $\lambda^c=\frac{\mu_a}{\Lambda_{\mbox{max}}(L)}\leq\frac{\mu_a}{\Lambda_{\mbox{max}}(A)}=\lambda_a^*$. For the same reason, one has $\lambda^c\leq\lambda_b^*$, regardless of the values of the inter-layer infection rates $\lambda_{ab}$ and $\lambda_{ba}$.

(2) The case of $\mu_a\neq\mu_b$.
One  obtains the following equations from Eq.~(\ref{matrix11}):
\begin{eqnarray}\label{matrix}
\lambda^cA\rho^{A}-\mu_aI_N\rho^{A}+\lambda_{ba}C\rho^{B}=0,\notag \\
\lambda^cB\rho^{B}-\mu_bI_M\rho^{B}+\lambda_{ab}D\rho^{A}=0.
\end{eqnarray}

  Usually, the  inter-layer infection rate is much smaller than the internal infection rate \cite{Cozzo2013Contact}, since  the  inter-layer infection rate describes  spreading   from one specie to another \cite{Morens2004The}, which is always slower than spreading within one species.  Thus, one can make two assumptions as follows:
\begin{eqnarray}\label{asump}
\lambda_{ba}\ll \lambda_a,~
\lambda_{ab}\ll \lambda_b.
\end{eqnarray}

Therefore, in order to compare the global epidemic  thresholds $\lambda^c$ of the interconnected  network with $\lambda_a^* $ and $\lambda_b^*$ of the corresponding  isolated layers, assume that $\lambda_a^*$ is close to $\lambda_b^*$, and then  use the perturbation method  to analyze the thresholds of  isolate layers. The perturbed solutions to thresholds $\lambda_a^*$  and $\lambda_b^*$ and infection rates $\rho^{A*}$ and $\rho^{B*}$ of the isolated layers can be written as
\begin{eqnarray}\label{perturbation}
&\lambda^c=\lambda_a^*+\epsilon_1 \lambda_a^*+O(\epsilon_1^2),\notag\\
&\lambda^c=\lambda_b^*+\epsilon_2 \lambda_b^*+O(\epsilon_2^2),\notag\\
&\rho^A=\rho^{A*}+\epsilon_3\rho^{A*}+O(\epsilon_3^2)\notag,\\
&\rho^B=\rho^{B*}+\epsilon_4\rho^{B*}+O(\epsilon_4^2).
\end{eqnarray}
Inserting Eq. (\ref{perturbation}) into Eq. (\ref{matrix}), using Eq. (\ref{isotatematrix}) and neglecting second-order terms, one obtains:
\begin{eqnarray}\label{eq:drive}
\epsilon_1\lambda_a^*A\rho^{A*}+\lambda_{ba}(1+\epsilon_4)C\rho^{B*}=0,\notag \\
\epsilon_2\lambda_b^*B\rho^{B*}+\lambda_{ab}(1+\epsilon_3)D\rho^{A*}=0.
\end{eqnarray}
Since $|\epsilon_3|\ll 1$ and $|\epsilon_4 |\ll 1$, and the elements in $A$, $B$, $C$ and $D$ are zero or positive numbers, it reveals that
$\epsilon_1<0$ and $\epsilon_2<0$, so that $\lambda^c<\lambda_a^*$ and $\lambda^c<\lambda_b^*$.  One can thus conclude that the global epidemic threshold of the interconnected network is smaller than the  thresholds of the corresponding isolated layers.
\section{Numerical simulations}\label{simulation}
 In  simulations,  network layer $A$ consisting of 1000 nodes and $B$  consisting of 800 nodes are respectively generated. 

\subsection{For the randomly-correlated homogeneous network}\label{homogeneous_networks_simulation}
  The WS algorithm \cite{Watts1998} is used to generate a small-world model for each layer. Specifically, for  layer $A$, start  with a ring of $N=1000$ nodes, each connecting to its $k_a$ nearest neighbors via undirected links. For the network $B$, start  with a ring of $N=800$ nodes, each connecting to its $k_b$ nearest neighbors via undirected links. The rewiring probability for   links is $0.2$   within each layer.  Then,   randomly connect a pair of nodes from the two layers, until the inter-layer average degree becomes about $k_a/2$. Monte Carlo  simulations  on Eq. (\ref{spread}) with different internal average degrees  are carried out to obtain the epidemic thresholds.  The initial fraction of infected nodes is set to 0.02, and the values of recovering rate are $\mu_1=1$ and $\mu_2=1$.  The   inter-layer infection rates are $\lambda_{ab}=0.1$ and $\lambda_{ba}=0.1$.

\begin{figure}
\includegraphics[width=0.5\textwidth]{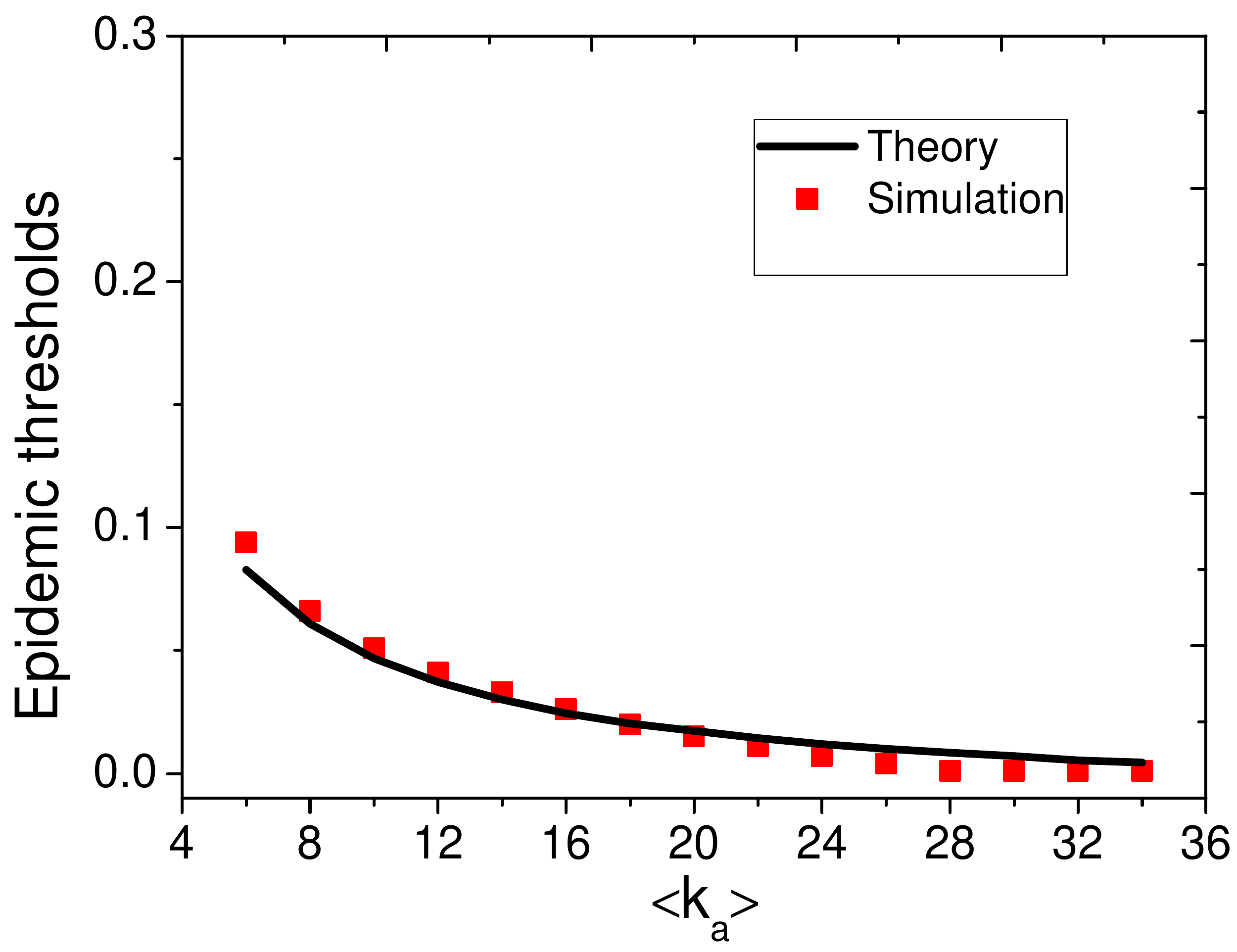}
\caption{ (Color online). Numerical simulation (red square) and theoretical (solid black line)  results of epidemic thresholds for the interconnected network  versus varying average degree $\langle k_a\rangle$.}
\label{Threshold_comparison_theory}
\end{figure}

  Let the internal average degree of $A$ be $k_a=(6,8,...,34)$, and  that of $B$ be  $k_b=(4,6,...,32)$.
The  comparison of the  global epidemic threshold from  theoretical analysis  described in Eq. (\ref{Seeking_roots})  with that from  numerical simulations  is displayed in Fig. \ref{Threshold_comparison_theory}. It shows that the theoretical analysis agrees  well with   numerical simulations with some minor deviation.

\begin{figure}
  \includegraphics[width=0.5\textwidth]{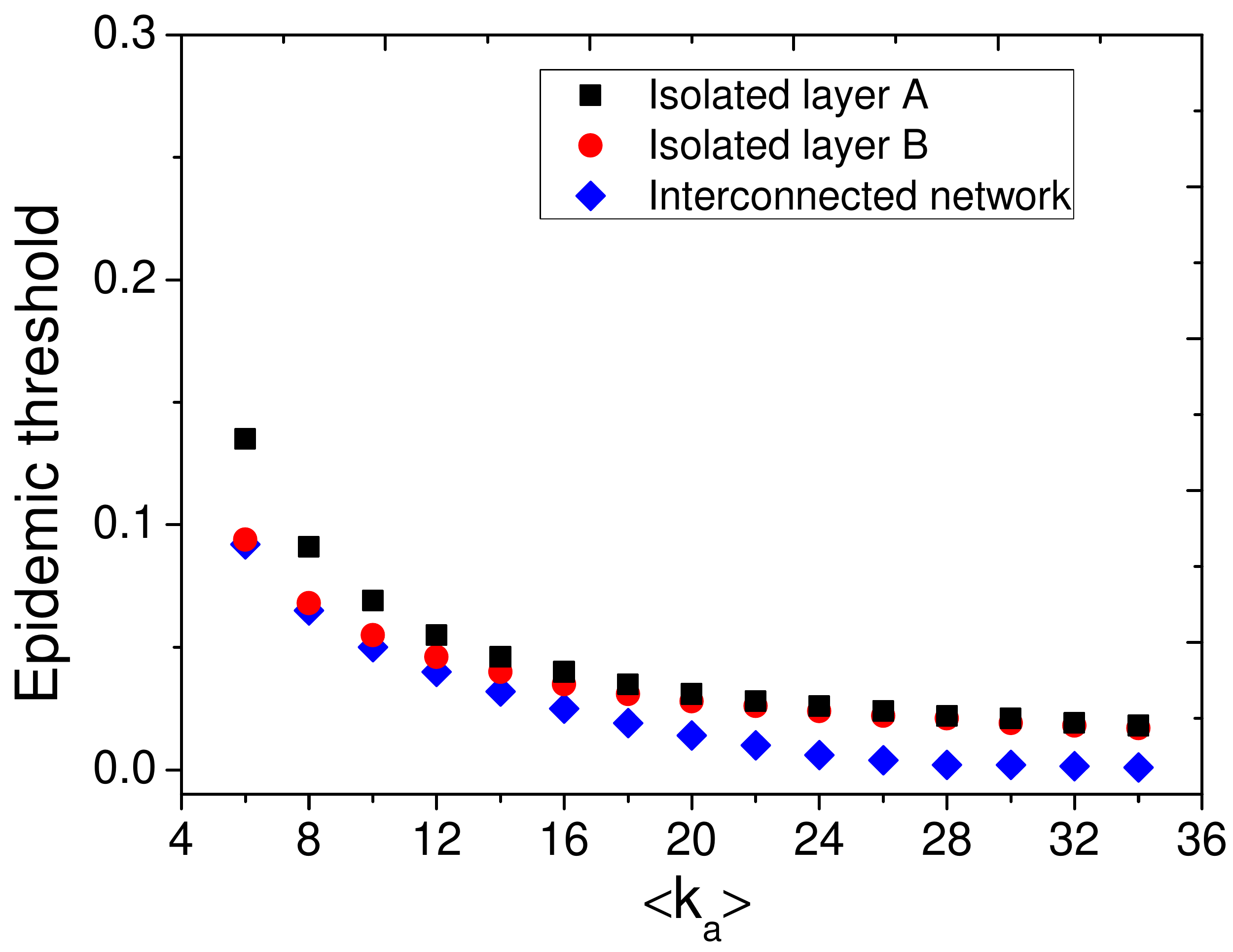}
\caption{ (Color online). Numerical epidemic thresholds  as a
function of $\langle k_a\rangle$  for the interconnected network and the corresponding isolated layers $A$ and $B$.}
\label{Threshold_comparison}
\end{figure}

The  comparison of the  global epidemic thresholds for the cooperative interconnected network with the thresholds of the  corresponding  isolated layers for different average degrees is displayed in Fig. \ref{Threshold_comparison}.  It shows that the global epidemic thresholds are always lower than those of the corresponding isolated layers, as indicated  by the  inequalities (\ref{global_threshold1}), (\ref{global_threshold2}), and (\ref{global_threshold3}). This observation verifies that cooperative epidemic spreading on an interconnected network promotes propagation.

\begin{figure}
  \includegraphics[width=0.5\textwidth]{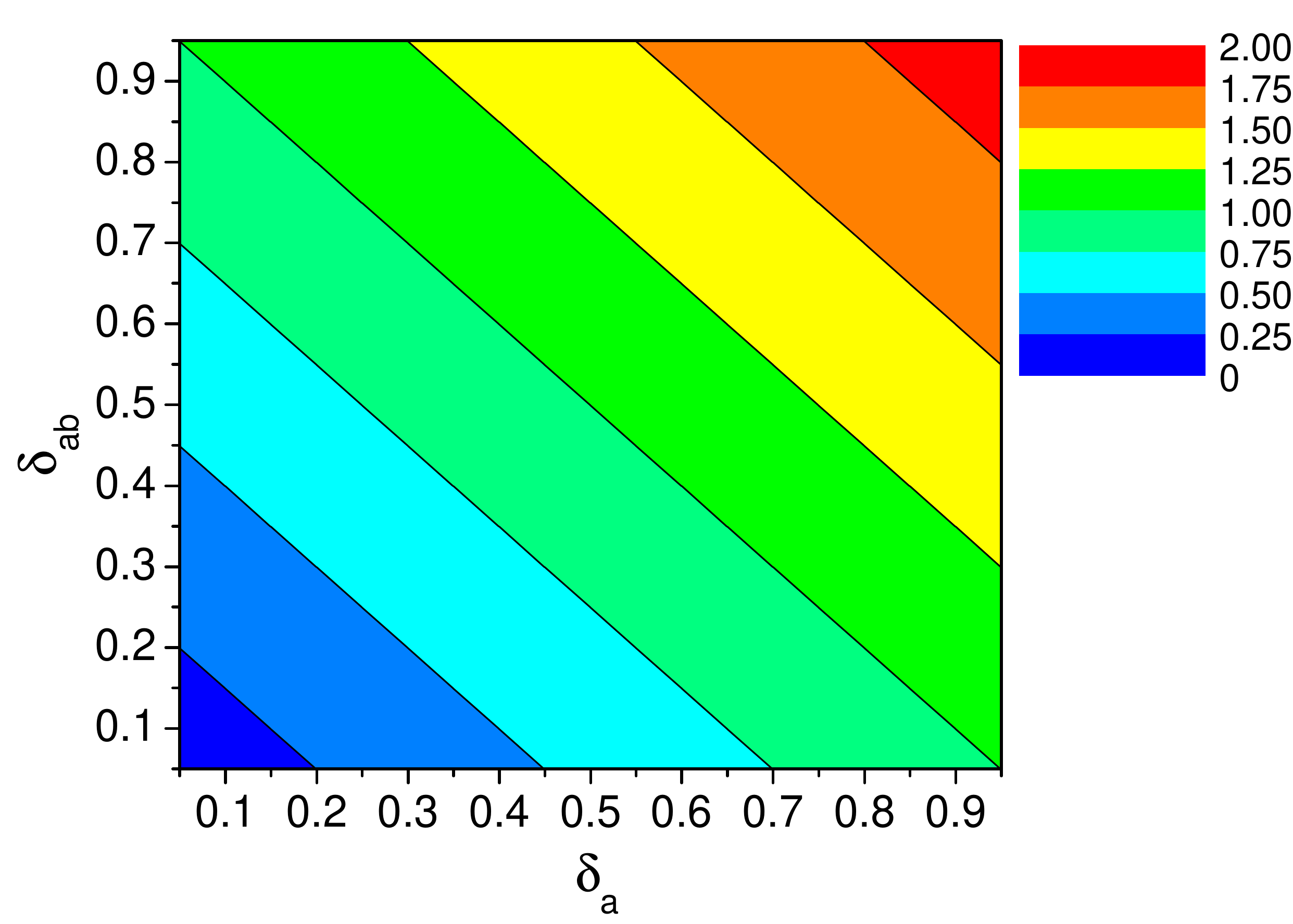}
\caption{ (Color online). Phase diagram of $\Lambda_{\mbox{max}}$  showing the healthy phase ($\Lambda_{\mbox{max}}<1$) and the
endemic phase ($\Lambda_{\mbox{max}}>1$) of the interconnected network for the case of $\delta_a=\delta_b$ and $\delta_{ab}=\delta_{ba}$.}
\label{Eigenvalues_greater_than_one1}
\end{figure}

\begin{figure}
  \includegraphics[width=0.5\textwidth]{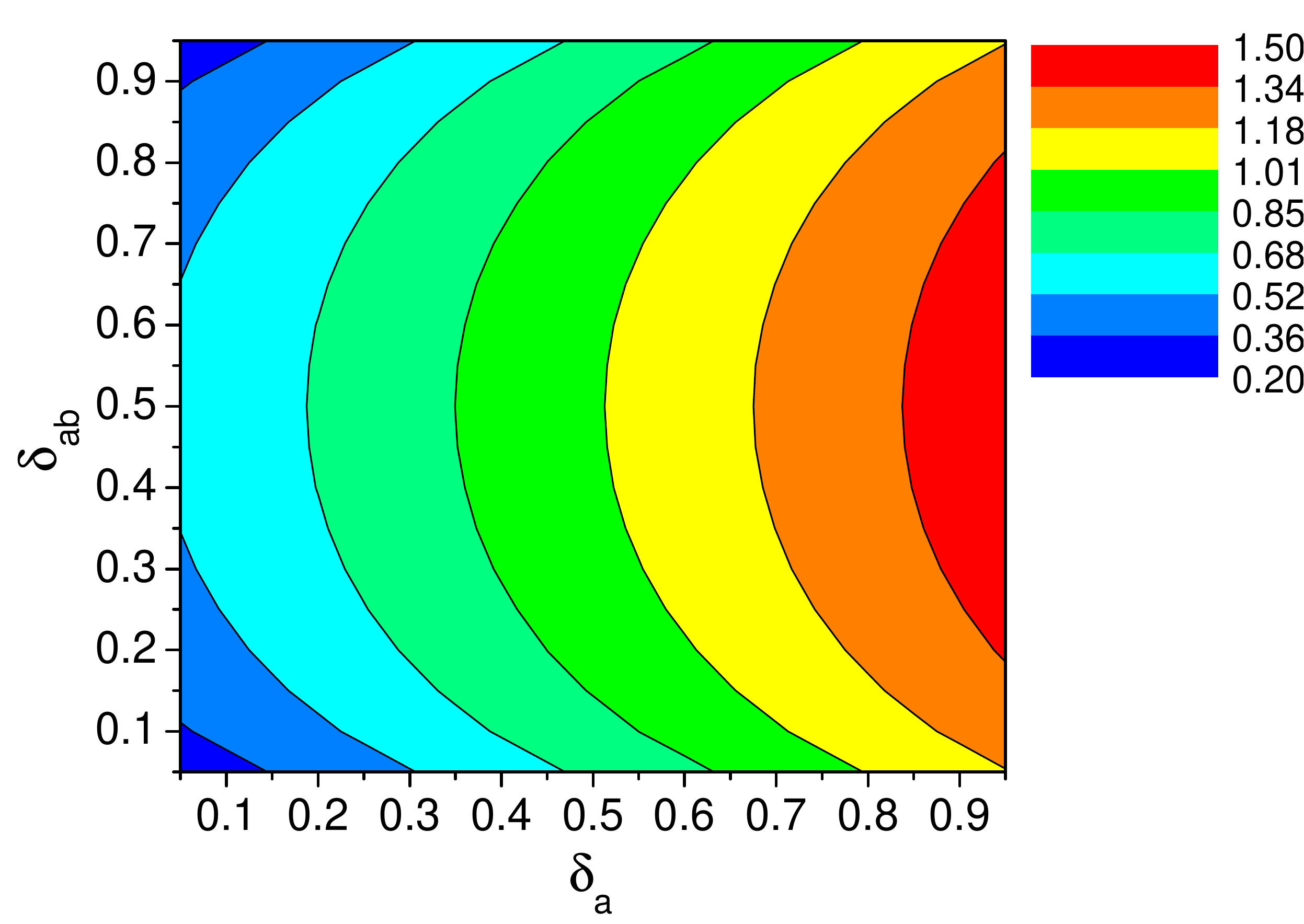}
\caption{ (Color online). Phase diagram showing the healthy phase~($\Lambda_{\mbox{max}}<1$) and the
endemic phase~($\Lambda_{\mbox{max}}>1$) of the interconnected network for the case of $\delta_a=\delta_b$ and $\delta_{ab}=1-\delta_{ba}$.}
\label{Eigenvalues_greater_than_one2}
\end{figure}

\begin{figure}
  \includegraphics[width=0.5\textwidth]{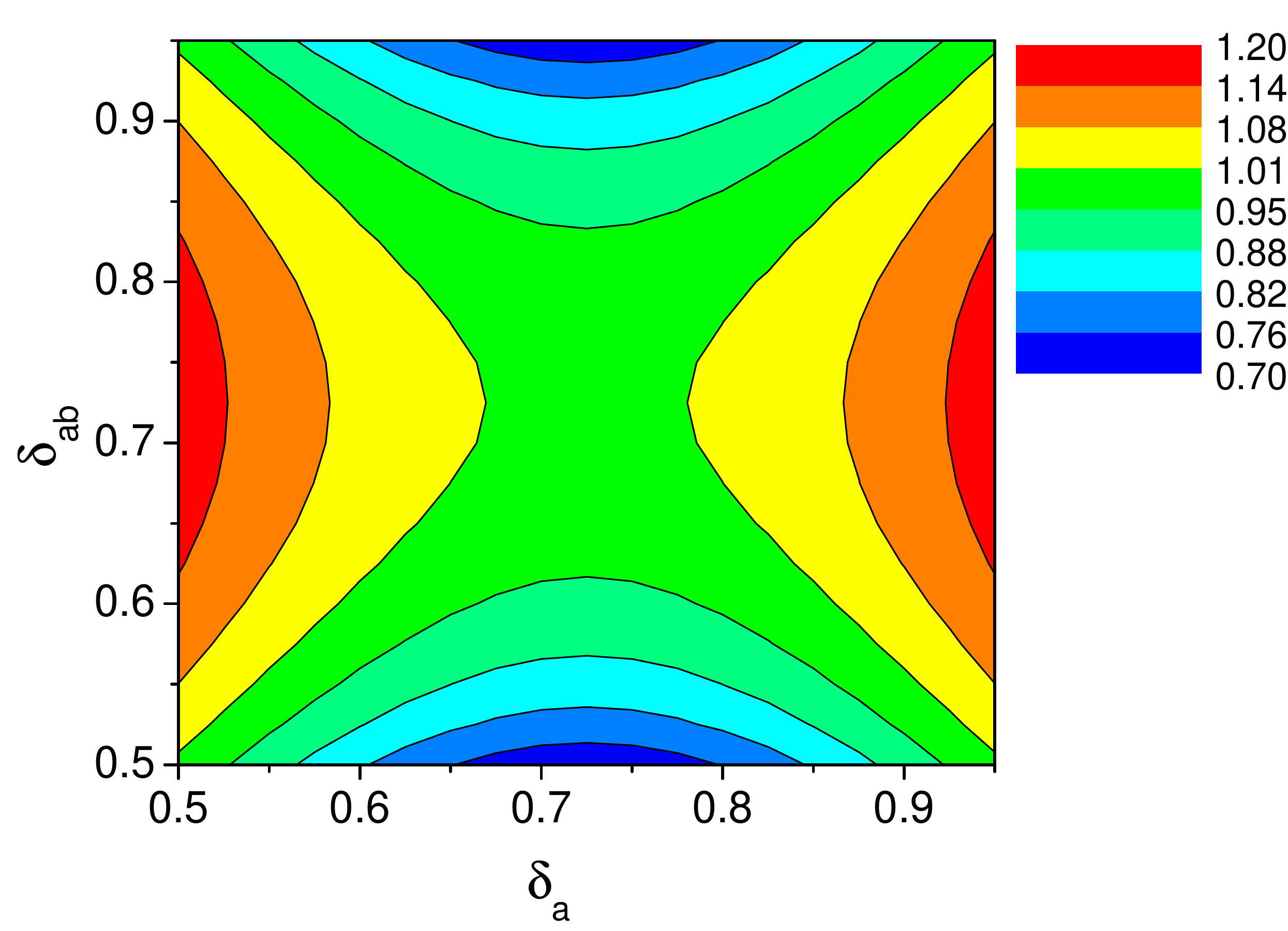}
\caption{ (Color online). Phase diagram of $\Lambda_{\mbox{max}}$  showing the healthy phase ($\Lambda_{\mbox{max}}<1$) and the
endemic phase ($\Lambda_{\mbox{max}}>1$) of the interconnected network for the case of $\delta_a=1-\delta_b$ and $\delta_{ab}=1-\delta_{ba}$.}
\label{Eigenvalues_greater_than_one3}
\end{figure}

Figures. \ref{Eigenvalues_greater_than_one1}, \ref{Eigenvalues_greater_than_one2} and \ref{Eigenvalues_greater_than_one3} show the phase diagrams for $\Lambda_{\mbox{max}}$ (the maximum eigenvalue of matrix $C$)  with respect to the parameters $\delta_a$ and   $\delta_{ab}$, for three specific cases.
Figure \ref{Eigenvalues_greater_than_one1} displays $\Lambda_{\mbox{max}}$    for the case of $\delta_a=\delta_b$ and $\delta_{ab}=\delta_{ba}$,  where the network evolves into an endemic phase  when $\Lambda_{\mbox{max}}>1$, and   it evolves into a healthy phase when $\Lambda_{\mbox{max}}<1$.  It is obvious that in the upper right triangular region regarding parameters $ (\delta_a,    \delta_{ab})$, an endemic state  arises, while in the other half part, the epidemic     dies out.

Figure \ref{Eigenvalues_greater_than_one2} shows  the phase diagram for the case of $\delta_a=\delta_b$ and $\delta_{ab}=1-\delta_{ba}$, where the endemic state arises  in the  right four regions, and the healthy state arises in the rest regions. Figure \ref{Eigenvalues_greater_than_one3} shows the phase diagram  for the case of $\delta_a=1-\delta_b$ and $\delta_{ab}=1-\delta_{ba}$, where the endemic state arises in the regions colored  in  green,  yellow, orange, and red. Furthermore, since $ \delta_a=\lambda_a\langle k_a\rangle, \delta_{ab}=
  \lambda_{ab}\langle k_{ab}\rangle $, when the average intra-  and inter-layer degrees are fixed,   whether the network is in an endemic phase or in a healthy phase is   determined by   the intra- and inter-layer infection rates.

  \begin{figure}
  \includegraphics[width=0.5\textwidth]{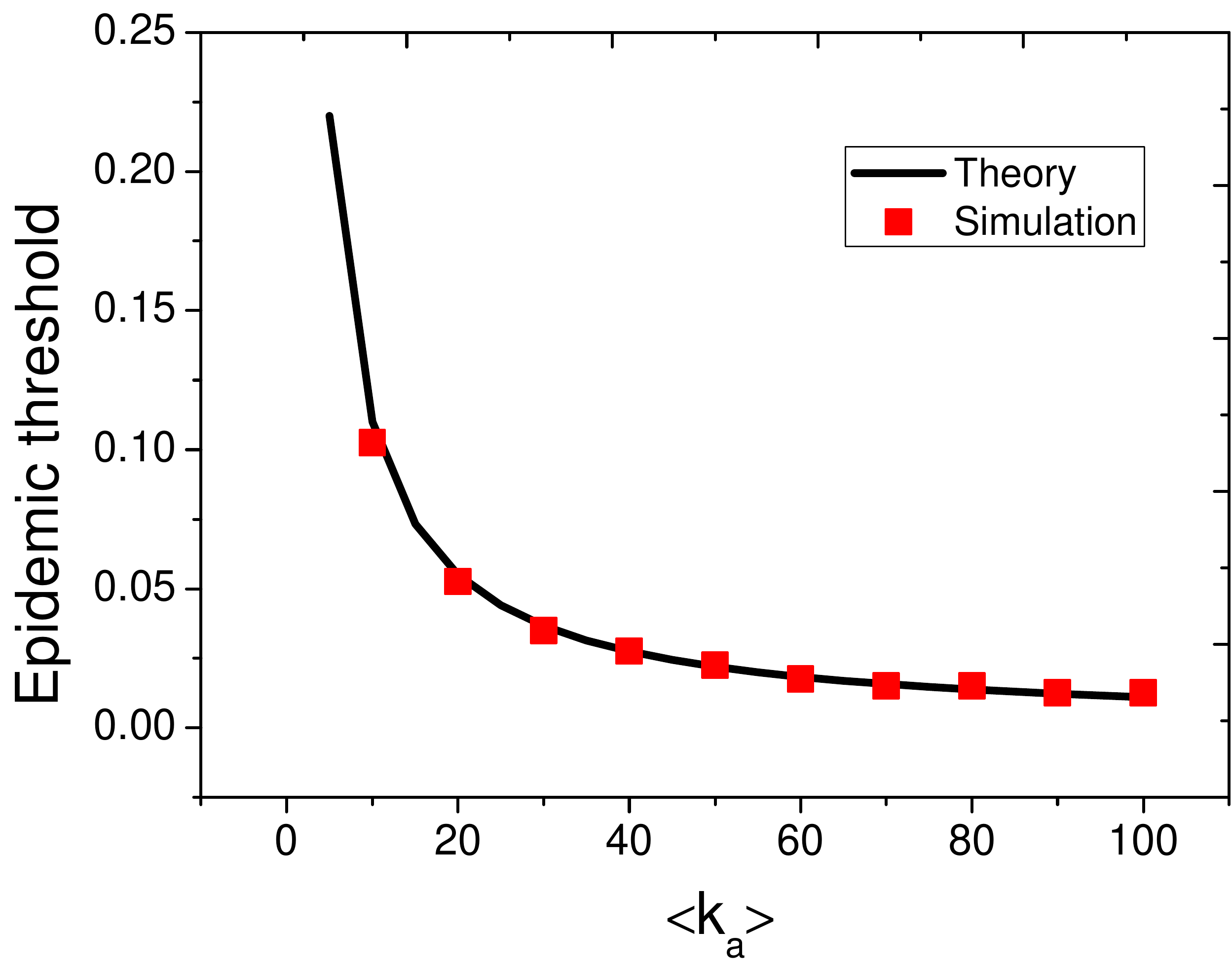}
\caption{ (Color online). Epidemic threshold of the interconnected network (red circum), along with the theoretical prediction (solid black line) based on different average degrees~$\langle k_a\rangle$.}
\label{homogenous_networks_1}
\end{figure}

 \begin{figure}
  \includegraphics[width=0.5\textwidth]{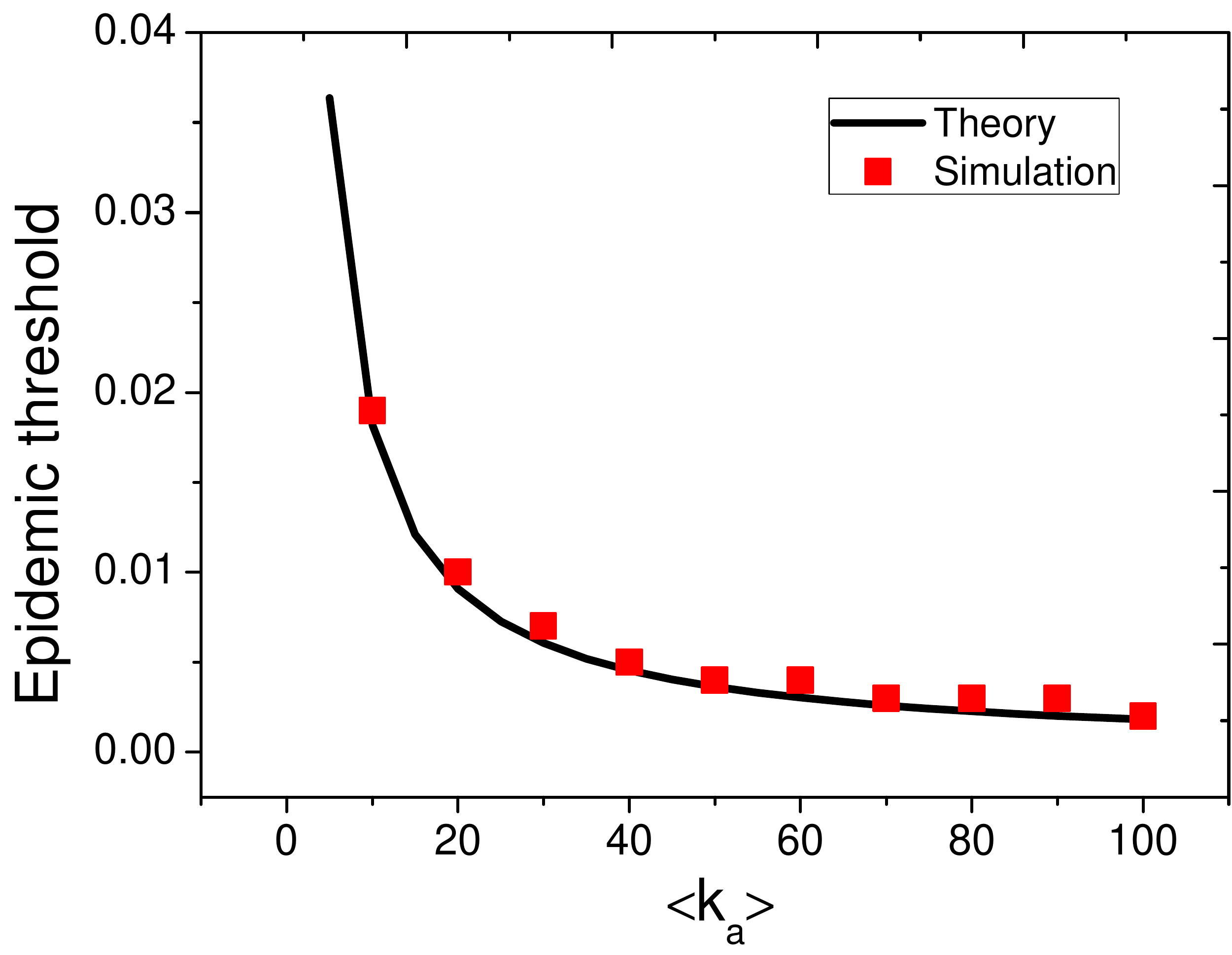}
\caption{ (Color online). Epidemic threshold of the interconnected network (red circum), along with the theoretical prediction (solid black line) based on different average degrees~$\langle k_a\rangle$.}
\label{homogenous_networks_2}
\end{figure}

 \begin{figure}
  \includegraphics[width=0.5\textwidth]{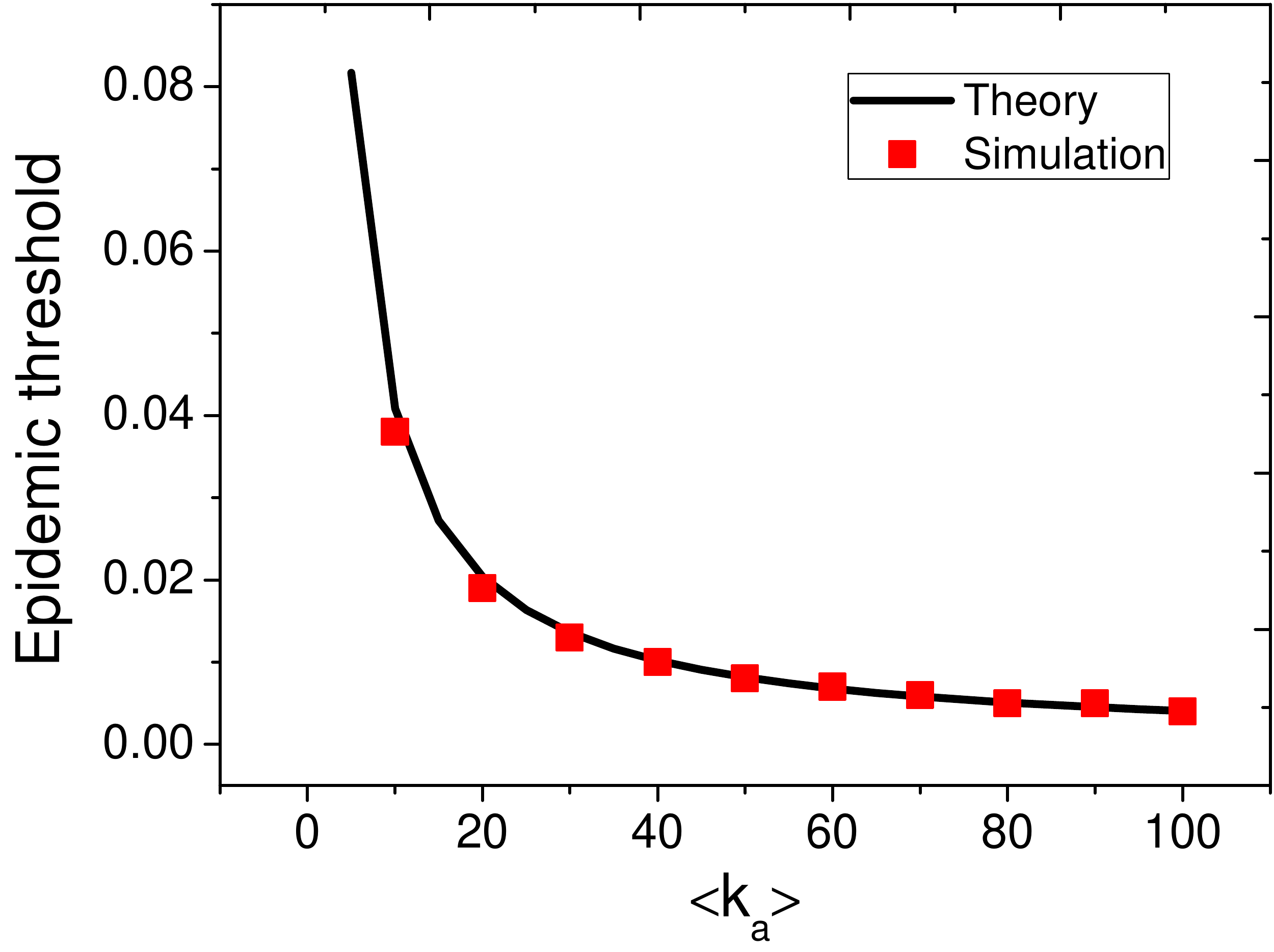}
\caption{ (Color online). Epidemic threshold of the interconnected network (red circum), along with the theoretical prediction (solid black line) based on different average degrees~$\langle k_a\rangle$.}
\label{homogenous_networks_3}
\end{figure}

Figures \ref{homogenous_networks_1}-\ref{homogenous_networks_3} verify Eqs. (\ref{E15})-(\ref{E17}) for the three specific scenarios.
In detail, Fig. \ref{homogenous_networks_1} shows the result for the case of $\lambda_{ab}\ll \lambda_a$,  $\langle k_{ab}\rangle <\langle k_{a}\rangle$ and $\langle k_{ba}\rangle<\langle k_{b}\rangle$).  It can be seen that the theoretical global epidemic threshold $\lambda^c=\frac 1{k_{max}}$  agrees well with   numerical simulations.

 Figure \ref{homogenous_networks_2} shows the result for  the scenario with $\lambda_{ab}\approx \lambda_a, \lambda_{ba}\approx \lambda_a$,  and $\langle k_{ab}\rangle \approx \langle k_{a}\rangle$, $\langle k_{ba}\rangle\approx\langle k_{b}\rangle$.  It can be seen that the
theoretical global epidemic threshold $\lambda^c=\frac 1{k_{a}+k_{b}}$ agrees well with numerical results.   Figure \ref{homogenous_networks_3} displays  the result for the scenario with $\lambda_{ab}\gg \lambda_a, \lambda_{ba}\gg \lambda_a$,  $\langle k_{ab}\rangle > \langle k_{a}\rangle$ and $\langle k_{ba}\rangle>\langle k_{b}\rangle$, which again shows that the theoretical result $\lambda^c=\frac 1{\sqrt{\langle k_{ab}\rangle \langle k_{ba}\rangle }}$ agrees well with numerical simulations.
 In the three figures,  one can see that   the epidemic threshold decreases sharply with increasing $\langle k_a\rangle$
when $\langle k_a\rangle$ is relatively small,  then the threshold keeps  declining but at a much smaller rate.

 \subsection{For the two-layered correlated heterogeneous network}\label{two-layered_correlated}
  \begin{figure}
  \includegraphics[width=0.5\textwidth]{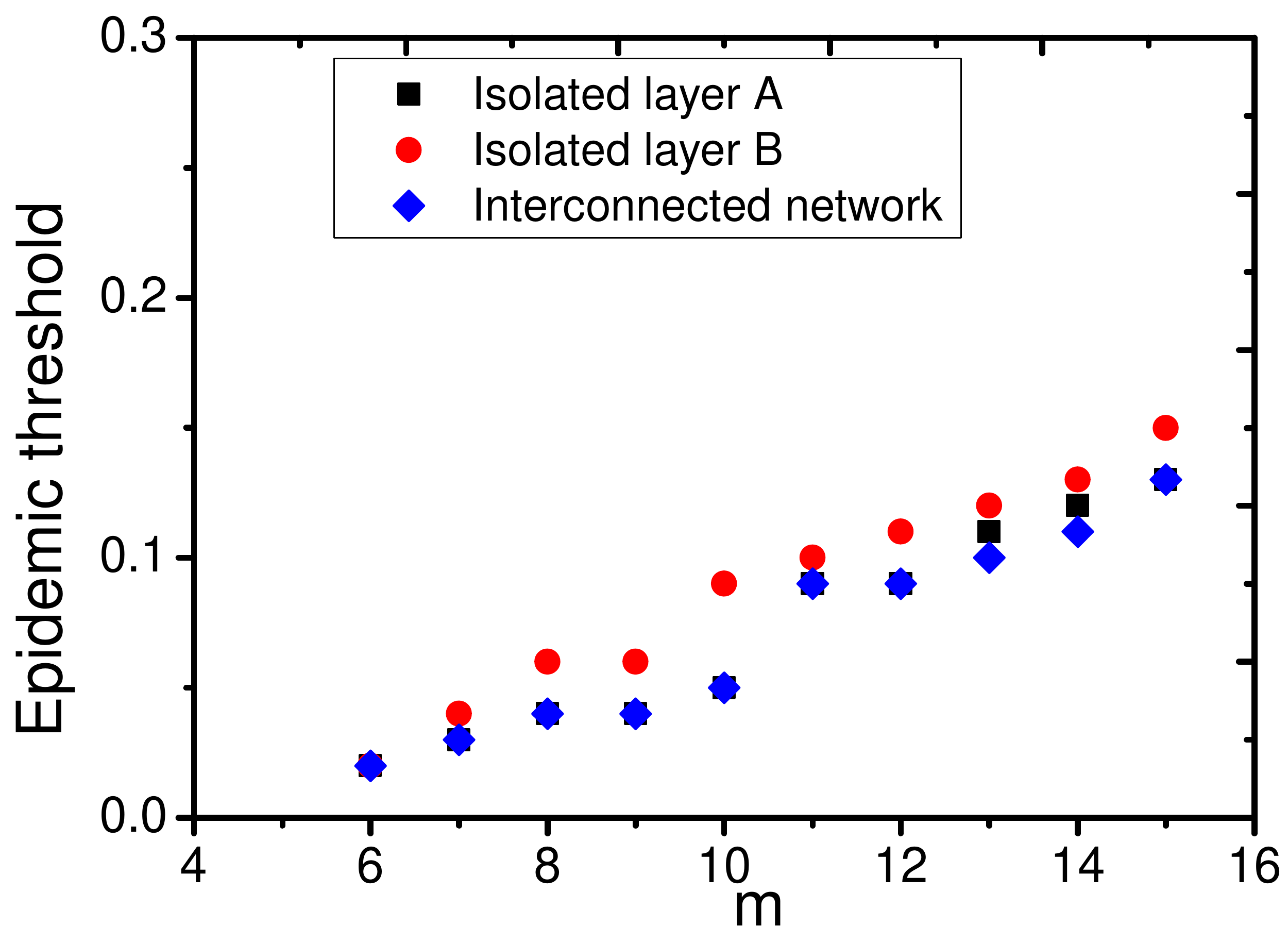}
\caption{ (Color online). {  Numerical epidemic thresholds  as a function of $m$  for two interconnected scale-free networks and the corresponding isolated layers $A$ and $B$ wherein.}}
\label{Comparison_ABInterconnected1}
\end{figure}

   \begin{figure}
  \includegraphics[width=0.48\textwidth]{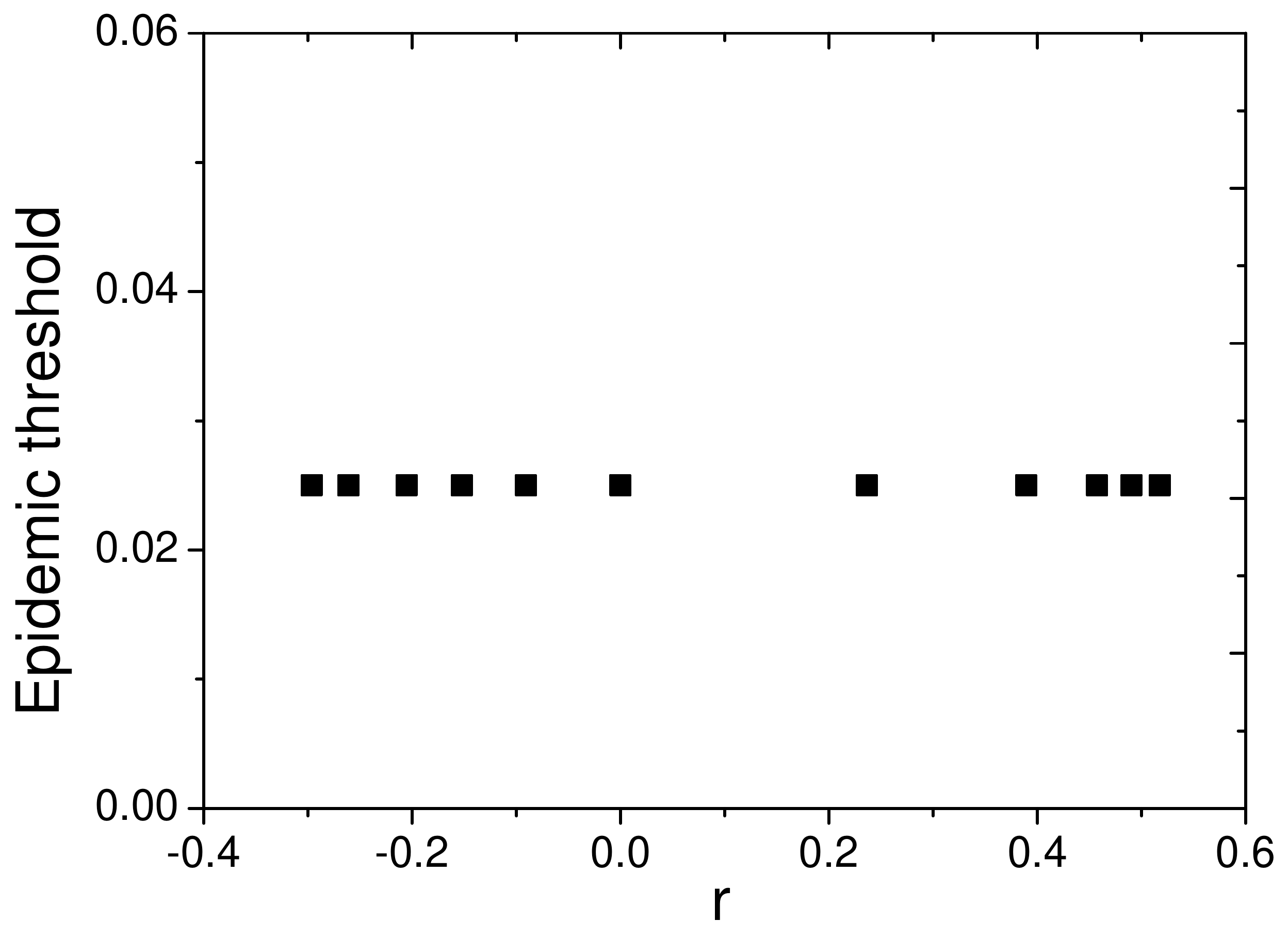}
\caption{ (Color online). The epidemic threshold  versus the  inter-layer degree correlation coefficient  $r$ for two interconnected scale-free networks.}
\label{Two_threshold}
\end{figure}
   \begin{figure}
  \includegraphics[width=0.48\textwidth]{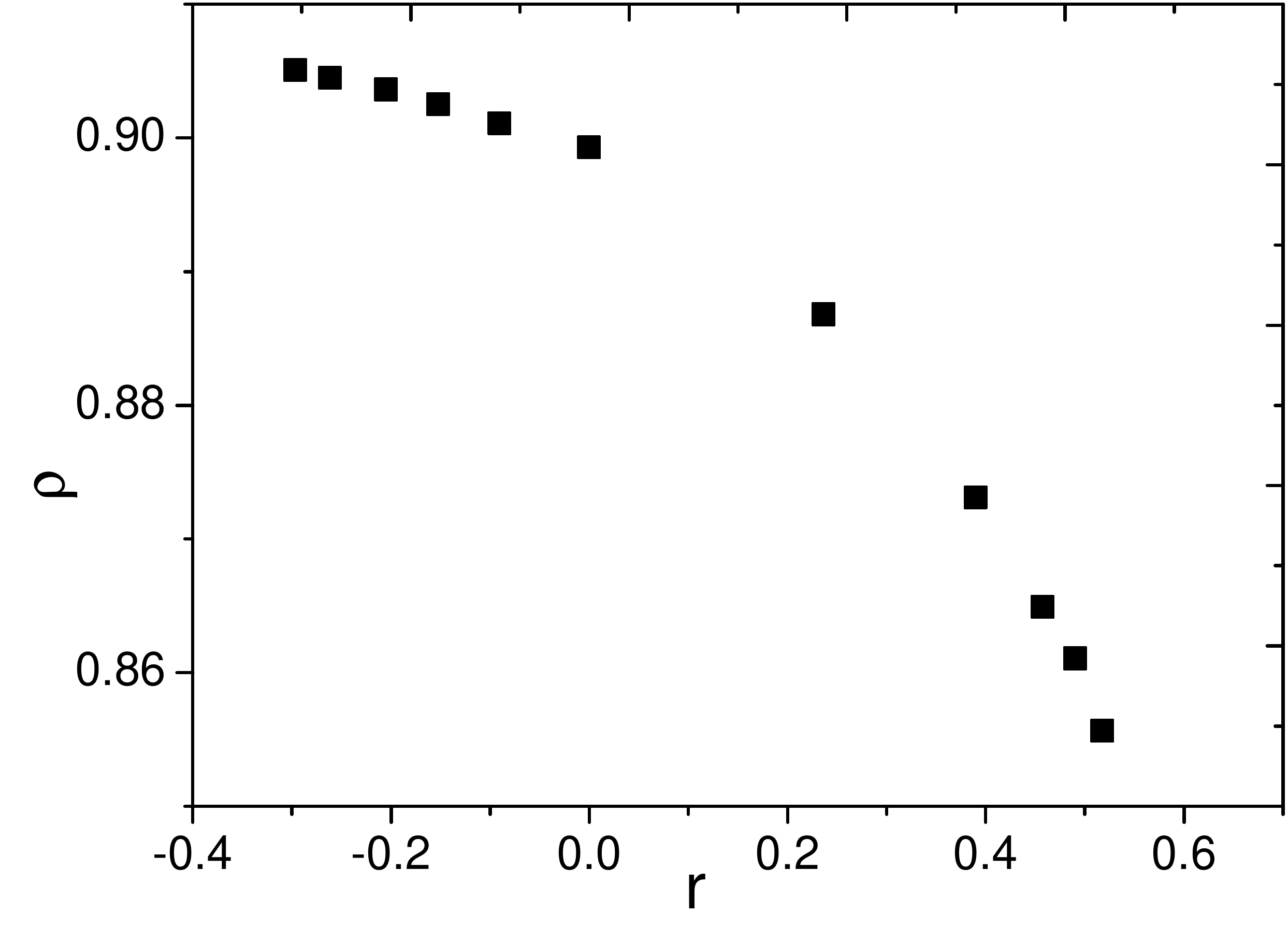}
\caption{ (Color online).  Total prevalence $\rho$  versus the inter-layer degree correlation coefficient  $r$ for two interconnected scale-free networks.}
\label{Two_Infection}
\end{figure}
 In this sub-section,   spreading processes on   interconnected BA scale-free networks \cite{Barabasi1999}  with inter-layer degree correlation are investigated.     
 The BA algorithm \cite{Barabasi1999} is employed to generate  networks for the two layers with identical model parameters but different sizes (i.e., 1000 for  layer $A$ and 800 for layer $B$ ).  Specifically, start with a  fully-connected network of $ m0=20 $   nodes.   At each time step,  add a new node,  which is connected to $m$ ($m$ is varying) existing nodes with a probability proportional to the number of links that the existing nodes already have.

  First, consider the  case where the correlation coefficient $r$ for nodes in the two layers  is $0$, which means that the two layers are uncorrelated.
   {

   In numerical simulations, {let  $m=(6,7,8,\cdots,15)$  for both layers $A$ and $B$.}   The  comparison of the  global epidemic threshold for cooperative interconnected network with the thresholds of the  corresponding  isolated layers for different scale-free networks are displayed in Fig. \ref{Comparison_ABInterconnected1}.  It shows that the global epidemic threshold is always lower than those of the corresponding isolated layers, which again verifies that cooperative epidemic spreading on an interconnected network promotes the propagation.


 Next, the problem of  how the degree correlation between two layers affects the disease spreading dynamics is investigated.  For the two scale-free networks, add  a fixed number of inter-layer connections (here, the number $\mathcal{  L }$ is half of the connections in layer $A$) but with adjustable values of correlation.
 Specifically, first, connect the  two  scale-free networks with a random  correlation, that is,  $r=0$.   Then, rewire $\lfloor \mathcal{  L } \delta \rfloor$ inter-layer links in such a way that the beginning end is kept and the other end is preferentially reconnected to another node bearing identical or nearly identical degrees so as to yield a larger degree correlation coefficient, where
  $\delta$ is the rewiring probability. Analogously, rewire $\lfloor \mathcal{  L } \delta \rfloor$ inter-layer links in such a way that the beginning end is kept and the other end is preferentially reconnected to another node bearing the most different degree to the beginning end so as to yield a smaller degree correlation coefficient.  Obviously,  different $\delta$ leads to different correlation coefficients.

  For two interconnected  $BA $ scale-free  networks both with $m=8$, the impact of inter-layer correlation $r$  on  the epidemic threshold is shown in Fig. \ref{Two_threshold}.  It can be seen that $r$ has little impact on the   thresholds.
     Further,  in Fig. \ref{Two_Infection}   the total prevalence $\rho$ as defined in Eq. (\ref{eq:rho26})   with varying  $r$ is shown for  the two scale-free networks, both with $m=8$. It is obvious that the prevalence decreases  with increasing $r$, which reveals that when the number of inter-layer connections is half of that in layer $A$,   a positive inter-layer node correlation will lead to a drop of total prevalence.


\subsection{For the uncorrelated two-layered  heterogeneous network}

Finally, to compare the epidemic spreading over isolated and interconnected networks, consider    two   BA scale-free networks, { where $m_0=10, m=10$ for layer $A$ and $m_0=8, m=8$ for layer $B$}.
Figure  \ref{3D_Isolate}   {shows  the total prevalence   $\rho^*$ defined in Eq. (\ref{prevalence2})} for the two isolated layers $A$ and $B$,  with respect to varying parameters $\lambda_a$ and $\lambda_b$.
 Figure \ref{3D_interconnected} shows {the total prevalence $\rho$ defined in Eq. (\ref{prevalence1})} of the interconnected network. {The blue regions ($\rho^*=0$ or $\rho=0 $) in the lower left  part in both figures show  that the epidemic  is  eventually dying out, while other regions ($\rho^*>0$ or $ \rho>0$) }indicate that the epidemic is eventually persistent in the population.  
The blue region in  Fig. \ref{3D_interconnected} for the interconnected  network  is much smaller than that in  Fig. \ref{3D_Isolate} for the isolated layers, which again reveals that epidemic threshold is decreased for cooperative epidemic spreading over the interconnected network.  Simultaneously, it can be seen from the different colors that for the same infection rates $\lambda_a$ and $\lambda_b$,  the prevalence for the
interconnected network is always larger than that for isolated layers.

 \begin{figure}
  \includegraphics[width=0.51\textwidth]{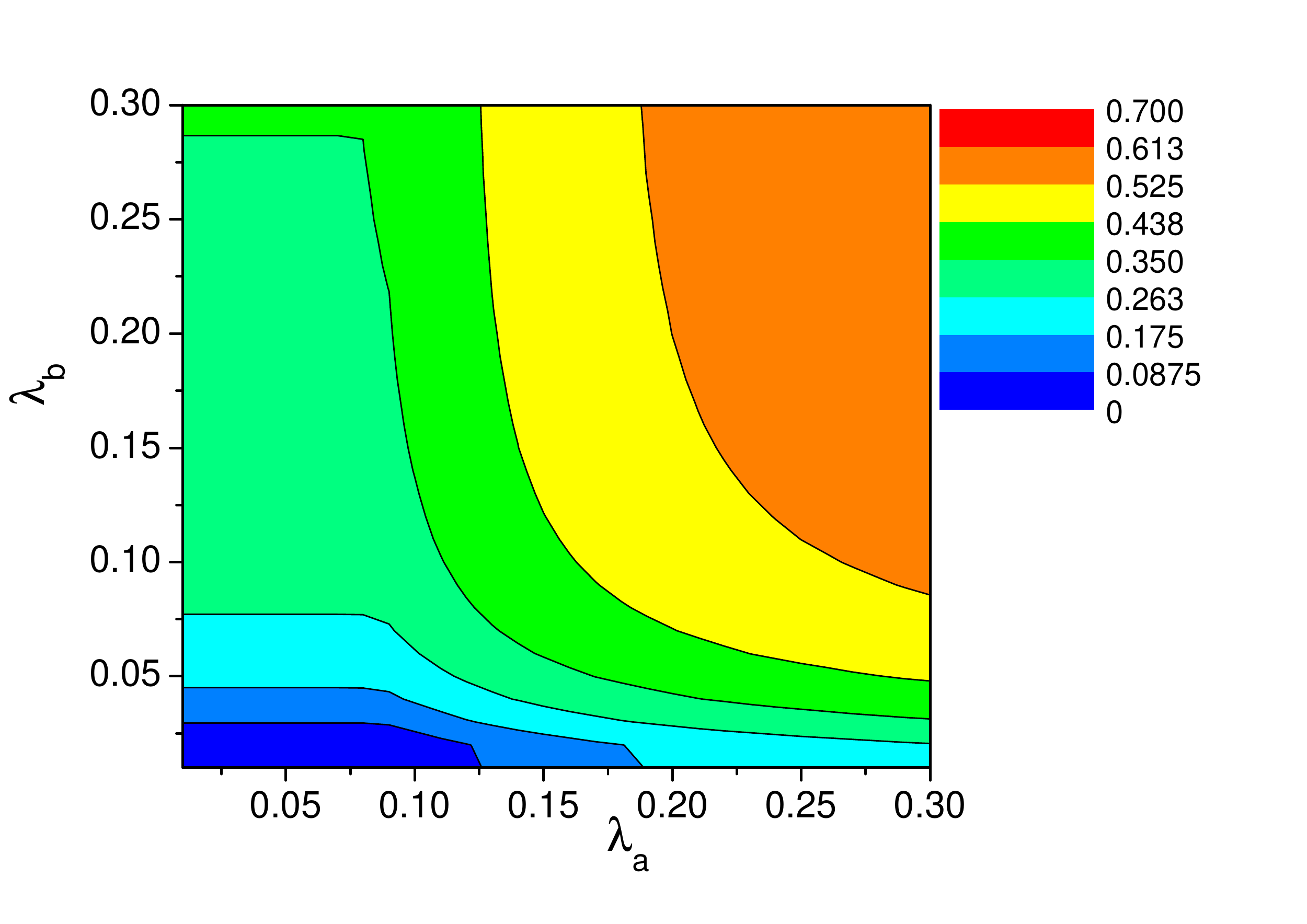}
\caption{ (Color online).  Total prevalence $\rho^*$ versus  infection rates $\lambda_a$ and $\lambda_b$ for the  two isolated layers.}
\label{3D_Isolate}
\end{figure}

 \begin{figure}
  \includegraphics[width=0.5\textwidth]{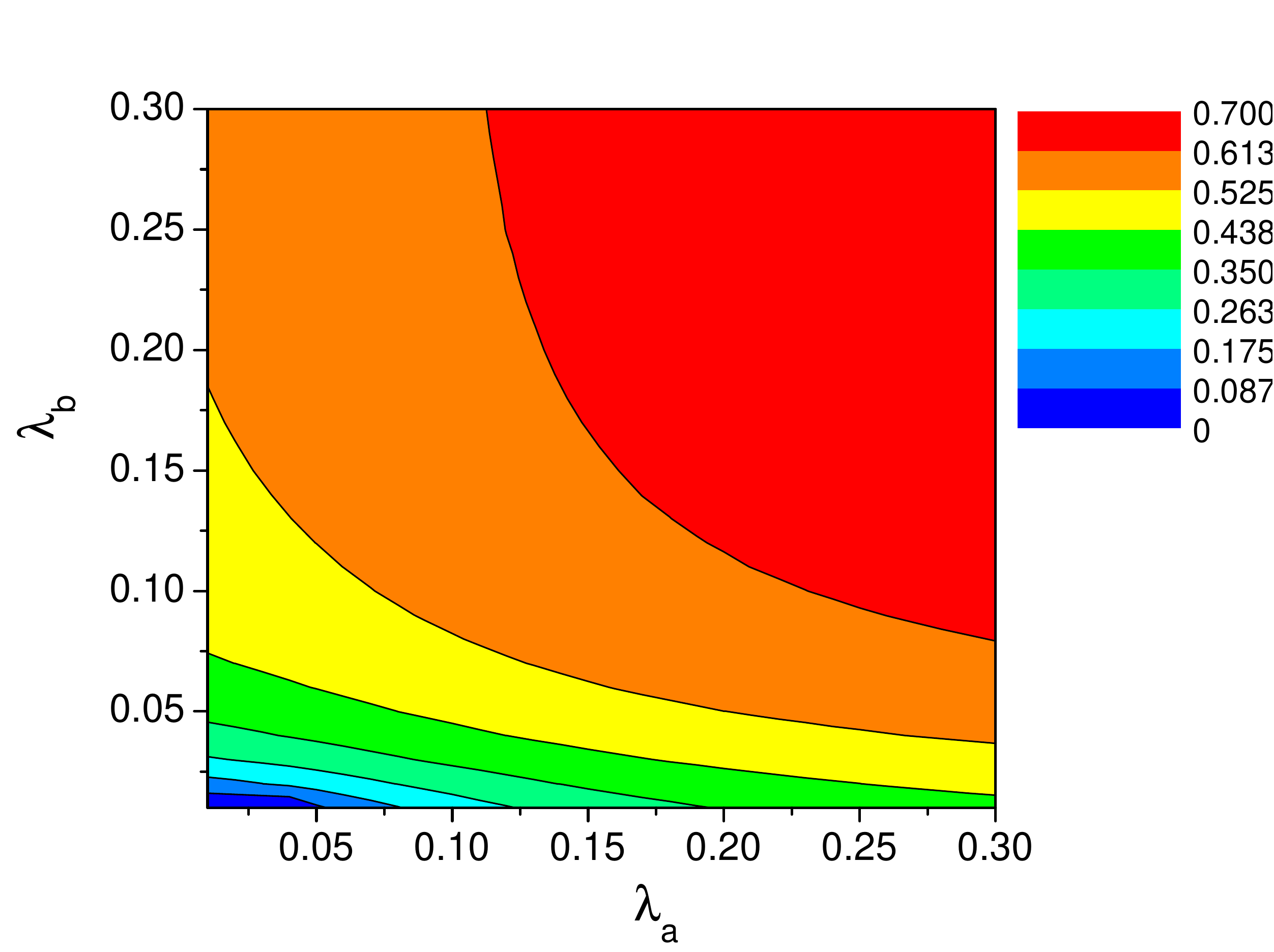}
\caption{ (Color online).  Total prevalence $\rho$ versus  infection rates $\lambda_a$ and $\lambda_b$ for the interconnected network.}
\label{3D_interconnected}
\end{figure}

\section{Conclusion}\label{Conclusion}
{
Three models have been formulated to investigate  cooperative  spreading processes on an
interconnected  network with or without inter-layer degree correlations.  In particular, for an interconnected homogeneous network, the dynamics has been theoretically analyzed at the level of each layer,  obtaining   global epidemic thresholds from information within each layer  and across layers.
 For  an interconnected heterogeneous   network with inter-layer correlations, it reveals that  inter-layer degree correlation  has little impact on the epidemic thresholds, but a larger inter-layer degree correlation coefficient  leads to a smaller total prevalence. {The global epidemic threshold is determined by the maximum eigenvalues of supra-connectivity matrix   and supra-adjacency marix for correlated and uncorrelated networks, respectively.}
 It was found that,   the epidemic thresholds of  spreading processes are decreased for interconnected networks, implying  that cooperative spreading processes  promote the spread of  diseases.  
 The results may provide  references to  public health monitoring  for disease control  and prevention. }

\nocite{*}
%

\end{document}